\documentclass[prd,aps,floats,floatfix,superscriptaddress,preprintnumbers,showpacs,eqsecnum,nofootinbib,twocolumn]{revtex4}
\usepackage{amsfonts,amsmath,epsfig,amssymb,latexsym,color,graphicx,array,subfigure,bm,
float }

\definecolor{red}{rgb}{1,0,0}
\definecolor{blue }{rgb}{0,0,1}
\definecolor{green}{rgb}{0,1,0}

\newcommand{\vect}[1]{\!\!\!\mbox{ \boldmath $#1$}}

\newcommand{\D}{{\rm d}}

\begin{document}
\title{Black Hole in the Expanding Universe from Intersecting Branes}
%---------------------------------------------------------------------%

\author{Kei-ichi {\sc Maeda}}
\email{maeda@waseda.jp}
\address{Department of Physics, Waseda University, 
Okubo 3-4-1, Shinjuku, Tokyo 169-8555, Japan}
\address{Waseda Research Institute for Science and Engineering,
Okubo 3-4-1, Shinjuku, Tokyo 169-8555, Japan}
\author{Masato {\sc Nozawa}}
\email{nozawa@gravity.phys.waseda.ac.jp}
\address{Department of Physics, Waseda University, 
Okubo 3-4-1, Shinjuku, Tokyo 169-8555, Japan}

%---------------------------------------------------------------------%
%--------------------------  Abstract  -------------------------------%
%---------------------------------------------------------------------%

\begin{abstract}
We study physical properties and global structures of a
time-dependent, spherically symmetric solution obtained  
via the dimensional reduction of intersecting M-branes.  
We find that the spacetime describes a maximally charged black hole
which asymptotically tends to the Friedmann-Lema\^itre-Robertson-Walker (FLRW)
universe filled by a stiff matter. The metric solves the field
equations of Einstein-Maxwell-dilaton system, in which four Abelian
gauge fields couple to the dilation with different coupling constants. 
The spacetime satisfies the dominant energy condition and 
is characterized by two parameters, $Q$ and $\tau$,
related to the Maxwell charge and the relative ratio of black-hole 
 horizon radii, respectively.  
In spite of the nontrivial time-dependence of the metric, 
it turns out that the black hole event horizon is a Killing horizon. 
This unexpected symmetry may be ascribed to the fact that the
11-dimensional (11D) brane configurations are supersymmetric in the
static limit. 
Finally, combining with laws of trapping horizon,
we discuss the thermodynamic properties of the black hole.  
It is shown that the horizon possesses a nonvanishing temperature, 
contrary to the extremal Reissner-Nordstr\"om (RN) solution.  
\end{abstract}

\pacs{
04.70.Bw, %Classical black holes 
%04.20.Dw,%Singularities and cosmic censorship 
%04.20.Ha,%Asymptotic structure 
04.50.+h,  %Gravity in more than four dimensions, Kaluza-Klein theory,
	  %unified field theories; alternative theories of gravity 
04.50.Gh  %Higher-dimensional black holes, black strings, and related objects
} 
\maketitle

%%%%%%%%%%%%%%%%%%%%%%%%%%%%%%%%%%%%%%%%%%%%%%%%%%%%%%%%%%%%%%%%%%%%%%%
%%%%%%%%%%%%%%%%%%%%%%%%%%%%%%%%%%%%%%%%%%%%%%%%%%%%%%%%%%%%%%%%%%%%%%%
\section{Introduction}
\label{sec:Introduction}
%%%%%%%%%%%%%%%%%%%%%%%%%%%%%%%%%%%%%%%%%%%%%%%%%%%%%%%%%%%%%%%%%%%%%%%
%%%%%%%%%%%%%%%%%%%%%%%%%%%%%%%%%%%%%%%%%%%%%%%%%%%%%%%%%%%%%%%%%%%%%%%

Brane configurations in string/M-theory have 
offered a new avenue for producing wide classes of 
solutions of physical interest in lower dimensions.  
The attempts of an early date have mainly aimed at the
construction of various kinds of black 
holes~\cite{GM,Strominger:1996sh,Maldacena:1996ky,Tanabe} 
via the Kaluza-Klein compactification of (intersecting) branes~\cite{
Gueven:1992hh,Argurio:1997gt,Ohta:1997wp,
Aref'eva:1997nz,Ohta:1997gw}.  
An interesting application of this idea is to 
dynamically realize our 4D universe by
incorporating time-dependence~\cite{Lu:1995cs,Lu2,Lu3}. 
An alternate mechanism that provides lower-dimensional spacetimes is 
the brane world~\cite{RS,BW}, in which our  
4D world is regarded as the hypersurface embedded in the bulk. 
Cosmological evolutionary scenarios based on the
dynamically moving brane or colliding branes  give a significant modification
from the standard cosmology in high energy r\'egime, still
consistent with the present day observations~\cite{
Kraus:1999it,Ida:1999ui,Khoury:2001wf}.

One can extend these studies further into the case where the brane involves  
a nontrivial space-and time-coordinate dependence.  Correspondingly, the 
4D reduced solution becomes spatially inhomogeneous and evolving in time. 
A preliminary result was presented in~\cite{Gibbons:2005rt}, 
where colliding D3 branes
were discussed within the framework of type IIB supergravity 
(see~\cite{Chen:2005jp} for analysis of the Ho\v{r}awa-Witten domain wall). 
Lower-dimensional solutions obtained by compactifying extended
directions of these moving branes or dynamically intersecting branes
have much richer properties than those obtained from static 
counterparts. 
Recently, the authors of~\cite{MOU} have studied  
dynamical solutions describing intersecting branes  
in more general settings and obtained a number of 
intriguing solutions with wide potential applications.  
Among other things, their tantalizing findings   
are the ``cosmological black holes''
in the expanding FLRW universe, that is,   
black holes  in a non-isolated 
system which asymptotically  approaches  
to the homogeneous and isotropic cosmology. 
In this paper, we are concerned with this 
``candidate'' black hole spacetime obtained in~\cite{MOU}.

Studies of black holes in our universe have been primarily focused upon the stationary
spacetimes in the literature~\cite{Carter}. Such approaches are definitely
 the first step because  we can anticipate that 
dynamic variations will die away and the system will settle down to
 equilibrium states if sufficiently long time passed after the formation of a
black hole. A number of physical properties of stationary black
holes  have been elucidated by many people.  
Specifically, black hole uniqueness theorem in stationary spacetime is the major 
triumph of mathematical relativity and  
established  that the Kerr solution describes all vacuum black holes
in isolated systems~\cite{Carter}. 
An essential crux toward the uniqueness  proof is the demonstration that the event horizon in 
stationary  spacetime is a Killing horizon~\cite{HE}. Since 
a Killing horizon is  a null surface to which the Killing vector is normal, we 
can identify locus of a black hole simply by the 
local spacetime symmetry. Furthermore, it has been revealed that Killing horizons admit three laws 
of black hole mechanics which bear an amazing 
resemblance to ordinary thermodynamics~\cite{
BHTD1,BHTD2,Hawking1974,Wald:1975kc,Wald:1993nt,Iyer:1994ys,Iyer:1995kg,
Gao:2001ut}. This 
implies a deep association between classical gravity,  statistical
physics and quantum mechanics, so that black hole thermodynamics is
expected to have a key r\^ole for understanding quantum aspects of gravity.
A notable progression in string theory is the microscopic 
derivation of black hole entropy in the perspective of 
intersecting brane configurations~\cite{Strominger:1996sh,Maldacena:1996ky}.

If we get rid of the stationarity assumptions to discuss dynamics, 
uniqueness theorem  no longer holds. Accordingly,  
a variety of black hole solutions are likely to exist. However, 
very little has been known concerning  the exact solutions of Einstein's
equations that describe growing black holes  
interplaying with surroundings.
A novel aspect of non-isolated system is that it
generally possesses a time-dependence and need not be asymptotically flat. 
The background fluid distributions filling the 
universe become inhomogeneous due to the presence of the black hole, 
while the black hole grows by swallowing the ambient matters.
Such a complexity has rendered the system
considerably elusive.

A large amount of effort has been devoted thus far to attempt to 
obtain black holes 
in the FLRW universe. An initiated work 
is the simplest model invented  by    
Einstein and Straus~\cite{Einstein:1945id}, which is often refereed to as a 
``Swiss-Cheese Universe.'' They matched 
 the Schwarzschild solution with an FLRW universe 
by means of a ``cut and paste'' method. So the black-hole part metric
still maintains a time-symmetry,  
and then it appears that this model does not capture realistic situation
of dynamic phase. 
If there is a positive cosmological constant,
we have the Schwarzschild-de Sitter (SdS) or
the Reissner-Nordstr\"om-de Sitter (RNdS) solution~\cite{SdS,Carter}.
Those spacetime can be redescribed by the coordinate transformation
in the form of a black hole in the exponentially expanding 
universe~\cite{BrillHayward}. 
However they have a ``timelike'' Killing vector and are essentially
static.

In recent years, 
Sultana and Dyer have constructed a more sophisticated black hole solution 
in a dynamical background by a conformal technique~\cite{SultanaDyer}. 
The matter content is composed of null dust and usual dust fluids, and
the solution tends to an Einstein-de Sitter spacetime asymptotically.  
This model, however, suffers from the issue of violating energy
conditions: energy densities of both fluids become negative at late times.

One of the other widely known black hole candidate in FLRW universe is the McVittie
solution~\cite{Mcvittie1933},  which is a spherically symmetric, 
expanding solution of the Einstein equations sourced by a perfect
fluid. Taking asymptotic limits,  the solution looks like an FLRW
universe at ``infinity,'' and like a black hole near the ``horizon,'' 
whence one might be led to conclude that the McVittie spacetime might 
describe a black hole in the expanding FLRW universe. 
Attractive as this might be,  however,  
such an optimistic outlook would jump to a hasty conclusion.     
As asserted in~\cite{Nolan}, the McVittie solution 
is disqualified as a black hole in an FLRW universe.
Since our spacetime metric 
is in appearance quite similar to the McVittie solution in several
respects,
the above concrete example motivates us to explore the global structure of 
time-dependent black holes with enough care. 
In this paper, we intend to provide a comprehensive account of the 
global picture of dynamical solutions obtained in~\cite{MOU}.

Another interesting issue of a time-dependent spacetime containing black holes 
is the collision of black holes.
Kastor and Traschen found the collection of extremely charged black holes
in the de Sitter universe and discussed their collision~\cite{KT,BHKT}.
This solution is a time-dependent generalization of its 
celebrated cousin, 
the Majumbdar-Papapetrou solution, which describes 
extremely charged RN black holes~\cite{Hartle:1972ya}. 
In spite of the lack of Bogomol'nyi-Prasad-Sommerfield
(BPS) states for $\Lambda >0$,  
it is somewhat astonishing that 
the superposition of RNdS black holes is possible.  
The same situation happens in our spacetime obtained 
by the time-dependent intersecting brane system
in which no BPS states are preserved. 
Making use of this exact solution, 
we can discuss the collision of black holes 
in the power-law contracting universe 
just as the brane collision~\cite{Gibbons:2005rt}.

The plan of the rest paper is as follows. We shall begin by 
reviewing the dynamical ``black hole solution'' derived in~\cite{MOU}. 
Section~\ref{sec:matter} involves a detailed examination 
about the properties of matter fields. Our main result is contained in  
Section~\ref{sec:spacetime_structure}, where we will elucidate
spacetime structures based on local and global perspectives. 
We draw the conformal diagram that allows us to pictorially 
identify a black hole
embedded in an expanding universe.  
Section~\ref{conclusion} summarizes our conclusive results
with several future outlook. In order to keep the mainstream 
of the text, we relegate some issues to Appendixes.

We shall work in the units of $c=\hbar=1$ and  retain the 
gravitational constant $\kappa^2=8\pi G$. 
We pertain to the notation $R$ to denote the circumference radius,
so that we use the script throughout the paper for the spacetime curvature 
${\cal R^\mu}_{\nu\rho\sigma}V^\nu:=2\nabla_{[\rho }\nabla_{\sigma ]}V^\mu$,
${\cal R}_{\mu \nu }:={\mathcal R^{\rho }}_{\mu \rho \nu }$
and 
${\cal R}:={\cal R^\mu }_\mu $.

%%%%%%%%%%%%%%%%%%%%%%%%%%%%%%%%%%%%%%%%%%%%%%%%%%%%%%%%%%%%%%%%%%%%%%%
%%%%%%%%%%%%%%%%%%%%%%%%%%%%%%%%%%%%%%%%%%%%%%%%%%%%%%%%%%%%%%%%%%%%%%%
\section{Time-dependent solutions from the intersecting brane system}
\label{sec:review}
%%%%%%%%%%%%%%%%%%%%%%%%%%%%%%%%%%%%%%%%%%%%%%%%%%%%%%%%%%%%%%%%%%%%%%%
%%%%%%%%%%%%%%%%%%%%%%%%%%%%%%%%%%%%%%%%%%%%%%%%%%%%%%%%%%%%%%%%%%%%%%%

The authors of~\cite{MOU} have classified the possible time-dependent
intersecting brane systems in M-theory by assuming the metric form,
and presented some interesting solutions in lower dimensions 
by compactification. Among other things, 
one of the most interesting solutions is a ``black hole''
in the expanding universe.
In the case where the all branes are at rest, i.e., 
spacetime is static, the  
4D reduced solution indeed describes a black hole, which is  obtained 
from M2-M2-M5-M5  brane system (four brane charges) or from M2-M5-W-KK
 brane system (two brane charges plus a Brinkmann wave and a Kaluza-Klein
monopole).
The 5D black hole solution is similarly derived
from M2-M2-M2  brane system (three brane charges)
or from  M2-M5-W  brane system (two branes plus a Brinkmann wave).
Time-dependent extensions of those lower-dimensional solutions
are produced from the dynamical intersecting brane systems,
in which only single brane is time-dependent~\cite{MOU}.
These 11D solutions and the procedure of the dimensional reduction  
are shortly summarized in Appendix~\ref{Intersecting_brane}.

After the toroidal compactification of 11D intersecting brane 
solutions, we have a 4D ``black hole solution,'' which is spherically symmetric, 
time and radial coordinate-dependent (we relegate the extension into the
case of non-spherical, multiple black holes to Appendix~\ref{multi_BH}).  
The 4D metric in the Einstein frame is given by
%--------------   4D_metric   ----------------%
\begin{eqnarray}
\D s^2=-\Xi \D t^2+{1\over \Xi}\left(\D r^2+r^2 \D \Omega_2^2\right)\,,
\label{4D_metric}
\end{eqnarray}
with 
\begin{eqnarray}
\Xi &=&
\nonumber 
\left[
\left(\frac{t}{t_0}+\frac{Q_T}{r}\right)
\left(1+\frac{Q_S}{r}\right)
\right.
\\
&&
\left.
~~
\left(1+\frac{Q_{S'}}{r}\right)
\left(1+\frac{Q_{S''}}{r}\right)
\right]^{-1/2}
\label{Xi}
\,,
\end{eqnarray}
where $\D \Omega_2^2=\D \theta^2+\sin^2\theta \D\phi ^2$
denotes the line element of a unit round two-sphere.
The constants $Q_T$ and $Q_S, Q_{S'}, Q_{S'\hspace{-.1em}'}$ are 
charges of one time-dependent brane and three static branes,
respectively.
Here and hereafter, the script ``$T$'' and ``$S$'' are understood to trace
their origin to time-dependent and static branes. 
The above metric manifests that 
the conditions of stationarity and asymptotic flatness were both relaxed.

Assuming $t/t_0>0 $ and changing to the 
new time slice $\bar t$ defined by
\begin{eqnarray}
{\bar t \over \bar t_0}
=\left ({t \over t_0} \right)^{3/4} ~~~~~{\rm with}~~~
\bar t_0=\frac{4}{3}t_0
\,,
\end{eqnarray}
we are able to put the solution (\ref{4D_metric}) 
into a more suggestive form, 
%--------------   4D_metric1   ----------------%
\begin{eqnarray}
\D s^2&=&-\bar \Xi \, \D \bar t^2+{a^2\over \bar \Xi}
\left(\D r^2+r^2 \D \Omega_2^2\right)\,,
\label{4D_metric1}
\end{eqnarray}
where 
\begin{eqnarray}
\bar \Xi &=&
\left[
\left(1+{Q_T\over a^4 r}\right)
\left(1+{Q_{S}\over r}\right)
\nonumber 
\right.
\\
&&
\left.
~~
\left(1+{Q_{S'}\over r}\right)
\left(1+{Q_{S'\hspace{-.1em}'}\over r}\right)\right]^{-1/2 }
\label{Xi1}
\,,
\end{eqnarray}
and 
\begin{eqnarray}
a &=&\left({\bar t\over \bar t_0}\right)^{1/3}
\label{scale_factor_4}
\,.
\end{eqnarray}

When we take the limit of $r\rightarrow \infty$, we can find that
the metric~(\ref{4D_metric1}) asymptotically tends to a flat FLRW spacetime, 
%----------------   FLW universe   ------------------%
\begin{eqnarray}
\D s^2&=&-\D \bar t^2+a^2
\left(\D r^2+r^2 \D \Omega_2^2\right)\,.
\end{eqnarray}
Here, the scale factor expands as $a\propto \bar t^{1/3}$, which is
the same as the expansion law of the universe filled by 
a stiff matter.
Hence we expect that this spacetime is asymptotically
FLRW universe with the equation of state $P=\rho$.

On the other hand, taking the limit  $r\rightarrow 0$ with $t$ being finite,
the time-dependence turns off and the metric~(\ref{4D_metric1}) reduces to 
the direct product of 2D anti-de Sitter (AdS$_2$) space with round
sphere, 
%------------- Near throat geometry  ----------%
\begin{eqnarray}
\D s^2&=&-{r^2\over \bar Q^2}\D t^2
+{\bar Q^2\over r^2}
\left(\D r^2+r^2\D \Omega^2_2\right)\,,
\label{near_horizon}
\end{eqnarray}
where $\bar Q:=(Q_TQ_{S}Q_{S'}Q_{S'\hspace{-.1em}'})^{1/4}$ plays the 
r\^ole of curvature radii of AdS$_2$ and S$^2$. 
This is a typical ``throat'' geometry of an extreme black hole.
To take the case of an extreme RN spacetime
with the Arnowitt-Deser-Misner (ADM) mass $M=\bar Q$, 
the metric in the isotropic coordinates reads 
%--------------  extreme RN  -------------------%
\begin{eqnarray}
\D s^2
&=&-\left(1+{\bar Q \over r}\right)^{-2}\D t^2
\nonumber \\
&&
+\left(1+{\bar Q \over r}\right)^{2}\left(\D r^2+r^2 \D \Omega^2_2\right)
\label{4D_RNBH}
\,,
\end{eqnarray}
which indeed asymptotes to the 
spacetime~(\ref{near_horizon}) in the limit $r\rightarrow 0$ 
with $t$ kept finite.\footnote{
It should be emphasized that the AdS$_2 \times$ S$^2$ 
geometry~(\ref{near_horizon}) indeed approximates the whole portion of the 
near-horizon geometry--the three dimensional null surface--of 
an extremal RN solution (and more generally, of 
any kinds of an extremal black hole~\cite{Kunduri:2007vf}),
not restricted to the ``throat.'' To the contrary, the
spacetime~(\ref{near_horizon}) fails to describe the near-horizon 
geometry of the event horizon of our time-dependent solution. 
As proof of the incident, we will show in later section that the
temperature of a dynamical black hole does not vanish in the present
case, unlike the extremal RN black hole 
having a zero Hawking temperature. 
}
Thus, we may speculate that there might exist a 
degenerate event horizon at $r=0$.

For this reason stated above, it might be tempted to regard the present 
spacetime~(\ref{4D_metric}) as a degenerate black hole with radius $\bar Q$ residing at the 
center $r=0$ of the expanding FLRW universe.
However, 
there is {\it a prioi} no guarantee that the 
global pictures of the solution is obtainable simply from the   
asymptotic considerations, just as we inferred the causal structure of 
the Schwarzschild-AdS spacetime from those of the Schwarzschild and  the
AdS metrics. Furthermore, 
it is far from clear to what extent the spacetime 
shares the physical properties as its limiting counterparts. 
In fact, a more detailed argument 
casts doubt on the above na\"ive expectation.

To see this, it is instructive to consider  
the McVittie spacetime~\cite{Mcvittie1933}, 
%-------------  McVittie spacetime  ----------------%
\begin{align}
\D s^2=&
-\left({1-{M/ 2ar}\over 1+{M/ 2ar}}\right)^2\D t^2
\nonumber \\
&+
a^2\left(1+{M\over 2ar}\right)^4\left(\D r^2+r^2 \D \Omega^2_2\right)
\,,
\label{McVittie}
\end{align} 
where $a=a(t)$. 
The metric (\ref{4D_metric1}) with (\ref{Xi1}) and~(\ref{scale_factor_4}) 
looks quite similar to the McVittie spacetime in appearance. 
In the limit of $r\rightarrow \infty$, the McVittie 
spacetime asymptotes to the flat FLRW universe with 
the scale factor $a$. If $a$ is set to be constant ($a\equiv 1$), 
we recover the Schwarzschild spacetime with
the ADM mass $M$ written in the isotropic coordinates. 
So one might deduce that the McVittie spacetime~(\ref{McVittie})
describes a black hole immersed in the FLRW universe.

However, 
it turns out that this exemplifies that our na\"ive estimate ceases to be 
true.\footnote{The case $a=e^{Ht}$ with constant $H$ is exceptional 
for which the metric describes an SdS spacetime. 
But we are reluctant to refer to this as a black hole in expanding
universe, 
since the metric is rewritten in a  static form thanks to Birkhoff's theorem.   
}
When the scale factor obeys the power-law form 
$a\propto t^p~ (p >0)$,  a curvature singularity appears at $r=M/(2a)$. 
As Nolan pointed out~\cite{Nolan},   
the spacetime events described by $r=M/(2a)$ in part consist of  
a shell-crossing spacelike singularity 
lying a future of a big bang singularity $a=0$. 
The surface $r=M/(2a)$ fails to describe a (regular) 
horizon of a black hole in FLRW universe. 
Inferring from the ``monopole term'' $m/r$, does 
this spacetime instead describe a point mass (singularity) at $r=0$
embedded in the expanding universe?
The answer is NO.  
It turns out that  $r=0$ corresponds to infinity,  
rather than the locus of a point particle in the universe. 
Besides that, 
according to the quasi-local definition of horizon, 
the McVittie spacetime may serve as a ``white hole'' in an expanding 
FLRW universe.  
As a good lesson of above,  
we are required to take special care to conclude what the present
 spacetime describes.

In this paper, we study the above spacetime~(\ref{4D_metric}) more thoroughly
[we are working mainly in Eq.~(\ref{4D_metric}) rather than Eq.~(\ref{4D_metric1}),
because the former coordinates cover wider range than the latter].   
We assume $t_0>0$, 
viz, the background universe is expanding.
For simplicity and definiteness of our argument, 
we will specialize to the case in which  all charges are equal, i.e.,
$Q_T=Q_{S}=Q_{S'}=Q_{S'\hspace{-.1em}'}\equiv Q~(>0)$.\footnote{
If $Q_T$ is different from other three same charges $Q_S$, 
the present result still holds. It is because 
such a difference amounts to the trivial conformal change  
$$
\D s^2=(Q_T/Q_S)^{1/2}\left[
-\Xi_* \D t_*^2+\Xi_*^{-1}(\D r^2+\D \Omega_2^2)\right]
$$  
with simple parameter redefinitions:
$\Xi_*=[(t_*/t_{*0}+Q_S/r)(1+Q_S/r)^3]^{-1/2}$,
 $t_*=(Q_T/Q_S)^{-1/2}t$,
and $t_{*0}=(Q_T/Q_S)^{1/2}t_0$.}
To be specific, we will analyze the spacetime metric 
%-------------- metric -----------------%
\begin{eqnarray}
\D s^2 =- \Xi\, \D t^2
+ \Xi^{-1}\left( \D r^2+ r^2 \D \Omega_2^2\right)
\label{metric}
\,,
\end{eqnarray}
whose component $\Xi $, Eq.~(\ref{Xi}), is simplified to 
\begin{eqnarray}
\Xi =\left(H_TH_S^3\right)^{-1/2}\,,
\end{eqnarray}
with
\begin{eqnarray}
H_T=\frac{t}{t_0}+\frac{Q}{r}, 
\qquad
H_S=1+\frac{Q}{r}
\label{metric_Xi}
\,.
\end{eqnarray}
A more general background with distinct charges are
yet to be investigated. 
The result for the 5D solution 
%(\ref{5D_metric}) 
will be given in Appendix~\ref{5DBH}.

%%%%%%%%%%%%%%%%%%%%%%%%%%%%%%%%%%%%%%%%%%%%%%%%%%%%%%%%%%%
\section{Matter fields and their properties}
\label{sec:matter}
%%%%%%%%%%%%%%%%%%%%%%%%%%%%%%%%%%%%%%%%%%%%%%%%%%%%%%%%%%%

It is a good starting point to
draw our attention toward the matter fields. 
Since we know explicitly the 4D metric components, we can read off 
the total energy-momentum tensor of matter fluid(s) from the
4D Einstein equations,
%-----------  Einstein's equations  ----------%
\begin{eqnarray}
\kappa^2T_{\mu\nu}=\mathcal G_{\mu\nu}
\,,
\end{eqnarray}
where 
$\mathcal G_{\mu\nu}=\mathcal R_{\mu\nu}-(\mathcal R/2)g_{\mu\nu}$
is the Einstein tensor.
What kind of matter fluids we expect?
There may appear at lest two fluid components:
one is a scalar field and the other is a U(1) gauge field.
This is because we compactify seven spaces
and we have originally 4-form field in 11D supergravity theory.
The torus compactification gives a set of scalar fields
 and the 4-form field behaves as a U(1) gauge field in 4D. 
In our solution, we assume four branes, which give rise to four
U(1) gauge fields.

As shown in Appendix~\ref{Intersecting_brane}, 
we can derive the following effective 4D 
action from 11D supergravity solution via compactification, 
%-------------- effective action  ----------------%
\begin{eqnarray}
{\cal S}&=&\int \D ^4x\sqrt{-g}\left[
{1\over 2\kappa^2}{\cal R}-{1\over 2}(\nabla\Phi)^2
\right.
\nonumber \\
&&~~~~~~~~~
\left.
-{1\over 16\pi}\sum_{A} e^{\lambda_A\kappa\Phi}
(F_{\mu\nu}^{(A)})^2\right]
\,,
\label{4D_action}
\end{eqnarray}
where $\Phi$, $F_{\mu\nu}^{(A)}$, and $\lambda_A$ 
($A=T,S,S',S'\hspace{-.1em}'$) are
a scalar field, four U(1) fields, and coupling constants, respectively.

%------------- Basic equations  ------------%
The above action yields the following set of basic equations, 
\begin{eqnarray}
&&
\mathcal G_{\mu\nu}=\kappa^2\left(T_{\mu\nu}^{(\Phi)}
+T_{\mu\nu}^{\rm (em)}\right)\,,
\label{4D_Einstein}
\\
&&
\Box \Phi -{\kappa \over 16\pi}\sum_{A} \lambda_A e^{\lambda_A\kappa\Phi}
(F_{\mu\nu}^{(A)})^2=0\,,
\label{eq_dilaton}
\\
&&
\nabla^\nu \left(e^{\lambda_A\kappa\Phi}
F_{\mu\nu}^{(A)}\right)=0
\label{eq_U(1)}
\,,
\end{eqnarray}
where
\begin{align}
T_{\mu\nu}^{(\Phi)}&=
\nabla_\mu \Phi\nabla_\nu \Phi -{1\over 2}g_{\mu\nu}(\nabla\Phi)^2\,,
\\
T_{\mu\nu}^{\rm (em)}
&=
{1\over 4\pi}\sum_{A}
e^{\lambda_A\kappa\Phi}\left[F_{\mu\rho}^{(A)}F_{~~\nu}^{(A)\rho}
-{1\over 4}g_{\mu\nu}(F_{\alpha\beta}^{(A)})^2\right]
\,.
\end{align}
For the present case with all the same charges, 
two different coupling constants
appear.

A simple calculation shows that 
our  spacetime metric~(\ref{metric}) satisfies 
the above basic equations (\ref{4D_Einstein}), (\ref{eq_dilaton}), and
(\ref{eq_U(1)}), 
provided  the dilaton profile 
%---------- dilaton ----------%
\begin{align}
\kappa\Phi&=
{\sqrt{6}\over 4}\ln\left({H_T \over H_S}\right)
\label{Phi}\,,
\end{align}
and four electric gauge-fields
%--------- Maxwell fields -----------%
\begin{align}
\kappa F_{01}^{(T)}
&=
-
{\sqrt{2\pi}Q\over r^2H_T^2}\,, 
\nonumber \\
\kappa F_{01}^{(S)}
&=\kappa F_{01}^{(S')}=
\kappa F_{01}^{(S'\hspace{-.1em}')}=
-
{\sqrt{2\pi}Q\over r^2H_S^2}
\label{form_field}
\,,
\end{align}
with the coupling constants  
\begin{align}
\lambda_T=\sqrt{6}\,,\qquad 
\lambda_{S}\equiv \lambda_{S'}\equiv
 \lambda_{S'\hspace{-.1em}'}=-\sqrt{6}/3. 
\end{align} 
%----------------------------------------%
The U(1) fields are expressed in terms of the electrostatic potentials 
$F_{\mu \nu }^{(A)}=\nabla_\mu A^{(A)}_\nu - \nabla_\nu A_\mu^{(A)}$ as, 
\begin{eqnarray}
\kappa A_{0}^{(T)}&=&{\sqrt{2\pi}\over H_T}\,,
\nonumber \\
\kappa A_{0}^{(S)}&=&\sqrt{2\pi}{\left({1\over H_S}-1\right)}
\label{U(1)_potential}
\,,
\end{eqnarray}
where we have tuned $A_0^{(S)}$ to assure  
 $A_0^{(S)}\rightarrow 0$ as $r\rightarrow\infty$ 
using a gauge freedom. 
Therefore the present spacetime~(\ref{metric})  is the exact solution
of the Einstein-Maxwell-dilaton system~(\ref{4D_action}).

One may verify that $Q$ is the (electric) charge satisfying
%----------- Maxwell charge ------------%
\begin{align}
\frac{Q}{\sqrt G}=\frac{1 }{4\pi} \int_S e^{\lambda_A\kappa \Phi}F_{\mu \nu }^{(A)}
\D S^{\mu \nu },
\end{align}
where $S$ is a round sphere surrounding the source.  
This expression is obtainable by the first integral of Eq.~(\ref{eq_U(1)}). 

Note that one can also find magnetically charged solution instead of
(\ref{form_field}). 
However, 
this can be realized by a duality transformation 
%------------- EM duality  --------------%
\begin{align}
\Phi \to -\Phi, \qquad
 F_{\mu \nu }^{(A)}\to \frac 12e^{\lambda_A\kappa \Phi}\epsilon_{\mu\nu\rho\sigma}F^{(A)\rho \sigma }\,,
\end{align}
which is a symmetry involved in the action~(\ref{4D_action}). 
Henceforth, we will make our attention only to the 
electrically charged case. This restriction does not affect the global
spacetime picture.

\subsection{Energy density and pressure}

Using our solution (\ref{4D_metric}), we can evaluate 
the components of the energy-momentum tensors, i.e., the energy density
and pressures for each field (
the dilaton $\Phi$ and U(1) fields $F_{\mu \nu }^{(A)}$). They are
given by 
%---------- Energy density and pressure  -------------%
\begin{widetext}
\begin{eqnarray}
\rho^{(\Phi)}=P^{(\Phi)}_r
={1\over 2}\left(\Xi^{-1}\,\dot{\Phi}^2+\Xi\,\Phi'^2\right)
&=&{3\over 16\kappa^2}\left[
{1\over t_0^2}\left({H_S\over H_T}\right)^{3/2}+
{Q^2\over r^4}{(H_T-H_S)^2\over (H_T^5H_S^7)^{1/2}}
\right] >0\,,
\label{Phi_energy}
\\
P^{(\Phi)}_\theta=P^{(\Phi)}_\phi
={1\over 2}\left(\Xi^{-1}\,\dot{\Phi}^2-\Xi\,\Phi'^2\right)
&=&{3\over 16\kappa^2}\left[
{1\over t_0^2}\left({H_S\over H_T}\right)^{3/2}-
{Q^2\over r^4}{(H_T-H_S)^2\over  (H_T^5H_S^7)^{1/2}}
\right]\,,
\label{Phi_pressure}
\\
\rho^{\rm (em)}=-P^{\rm (em)}_r=P^{\rm (em)}_\theta=P^{\rm (em)}_\phi
&=&
{1\over 8\pi}\left[
e^{\lambda_T\kappa\Phi}
(F_{01}^{(T)})^2
+3e^{\lambda_S\kappa\Phi}
(F_{01}^{(S)})^2
\right]\,,
\nonumber \\
&=&{Q^2\over 4\kappa^2r^4}\left[
{1\over H_T^4}\left({H_T\over H_S}\right)^{3/2}+
{3\over H_S^4}\left({H_T\over H_S}\right)^{-1/2}
\right]>0
\label{em_energy}
\,.
\end{eqnarray}
\end{widetext}
The time $t=t_0$ is special at which 
the energy density of 
the scalar field is uniform 
$\kappa^2 \rho^{(\Phi)} (t_0)=3/(16t_0^2)$, 
and soon after it becomes gradually inhomogeneous. 
We shall refer to $t_0$ as the {\it fiducial time}.
As the universe expands, the energy density of the scalar field
at infinity behaves
$\rho^{(\Phi)}\propto t^{-3/2}\propto \bar t^{-2}\propto a^{-6}$,
as expected for the FLRW universe with stiff matter
or a massless scalar field. 
The energy density of the U(1) fields evaluated at the 
fiducial time is 
$\kappa^2\rho^{\rm (em)} (t_0)=Q^2/(rH_S)^4=Q^2/(r+Q)^4$, which is 
the same as that of the extreme RN spacetime,
and decreases in time as 
$\kappa^2\rho^{\rm (em)}(t, r)\propto t^{-1/2}\propto\bar t^{-2/3}\propto a^{-2}$ 
near infinity as in the same manner for the FLRW universe.

%-------------  energy flux  ---------------%
We can also find the energy flux $J^\mu :=T^{\hat 0 \mu }$ whose 
spatial component ${\cal F}$ is given by 
\begin{align}
\kappa^2{\cal F}
=\kappa^2 T^{\hat 0}_{~\hat 1}=\kappa^2 {T^{(\Phi)\hat 0}}_{\hat 1}
=-{3Q\over {8}t_0} {(H_T-H_S)\over r^2 H_T^2 H_S}
\,,
\label{flux}
\end{align}
where a hat denotes the tetrad component.
Only the scalar field contributes to the energy flux, 
since there exist no magnetic fields, i.e., no Poynting flux. 
One can find  that the $\mathcal F$ becomes negative for $t>t_0$, implying that
the scalar field energy is falling  toward the black hole.
However, as it turns out later, the flux never gets into the black hole.
This may be attributed to that 
the repulsive force caused by the U(1) fields 
becomes strong near the horizon and
finely balances the attractive gravitational force 
of the dilaton field.

It is worthwhile here to discuss the issue of the energy conditions.
Many candidates of  ``black hole solutions'' in 
expanding universe found in the literature
(McVittie's solution~\cite{Mcvittie1933}, 
Sultana-Dyer solution~\cite{SultanaDyer}, etc)
do not respect
energy conditions in the whole of spacetime. 
Now the present system--U$(1)$ gauge fields coupled to
dilaton--apparently satisfies the energy condition and hence
it provides us a nontrivial example. 
We may verify this explicitly as follows.
Inspecting Eqs. (\ref{Phi_energy})--(\ref{em_energy}), 
we notice  
$\rho^{(\Phi)}= P_r^{(\Phi)} \ge P_\theta^{(\Phi)}$ and
$\rho ^{(\rm em)}=-P_r^{(\rm em)} \ge P_r^{(\rm em)}$, 
from which we obtain
\begin{align}
\rho =& \rho^{(\Phi)}+\rho^{\rm (em)}
=P_r^{(\Phi)}-P_r^{(\rm
 em)}\nonumber \\
\ge & P_r^{(\Phi)}+P_r^{(\rm em)}=P_r,
\nonumber \\
\rho =&\rho^{(\Phi)}+\rho^{\rm (em)}\ge 
P_\theta^{(\Phi)}+P_\theta^{(\rm em)}=P_\theta
\,,
\end{align}
and
\begin{align}
 \rho+P_r&= (
\rho^{(\Phi)}+P_r^{(\Phi)}
)+(\rho^{(\rm em)}+P_r^{(\rm em)})>0\,, \nonumber \\
\rho+P_\theta &=(\rho^{(\Phi)}+P_\theta^{(\Phi)})
+(\rho^{(\rm em)}+P_\theta^{(\rm em)})>0\,.
\end{align}
These equations mean that $\rho\ge |P_i| ~(i=1,2,3)$ is
satisfied anywhere.

The energy flux  $J^\mu =T^{\hat 0\mu }$
satisfies
\begin{align}
J^\mu J_\mu &=-\rho^2+\mathcal F^2 \nonumber \\
&=-(\rho^{({\rm em })})^2
-2\rho^{({\rm em})}\rho^{(\Phi)}-(P_\theta^{(\Phi)})^2
<0\,,
\end{align}
where we have used the relation
$(\rho^{(\Phi)})^2-\mathcal F^2=(P_\theta^{(\Phi)})^2$
at the second equality.
Hence, the energy current $J^\mu $ 
is a timelike vector everywhere. 
It then follows that  the spacetime~(\ref{4D_metric}) 
satisfies the dominant energy condition 
[$\rho\ge |P_i| ~(i=1,2,3)$ and $J^\mu $ is non-spacelike].

\bigskip

\subsection{Misner-Sharp energy}

Another useful quantity to characterize matter fields is the 
quasilocal energy, which is defined on the closed two-surfaces. 
If the spacetime has an SO($3$)-symmetry, we are able to
give a physically satisfactory quasilocal 
energy introduced by Misner and Sharp~\cite{MS1964}.

The utility of spherical symmetry lies in the fact that we
can covariantly employ the circumference radius,  
\begin{align}
R(t, r):= |r| \Xi^{-1/2}=|r|(H_TH_S^3)^{1/4},
\label{R}
\end{align}
in terms of which the area of metric sphere is given by 
$4\pi R^2$. $R$ is a geometrical quantity 
and has an invariant meaning.  
It is occasionally of great advantage,  instead 
of the comoving coordinate $r$, to make use of $R$. 
Using the circumference radius, 
 the Misner-Sharp quasilocal energy is defined by~\cite{MS1964},
\begin{align}
 m(t,r) 
:=\frac{4\pi R}{\kappa^2} \left[1-g^{\mu \nu }(\nabla_\mu  R)(\nabla_\nu R)\right]\,.
\end{align}
This quantity is a useful local measure to demonstrate geometric properties of
spacetime~\cite{Hayward1994,hideki,nozawa}. The Misner-Sharp energy 
represents a mass energy contained inside the surface of radius $R$. 
Once the compact surface is specified, the Misner-Sharp mass is given 
without any ambiguity. 
Such a quasi-localization is possible because of spherical symmetry, in which no  
gravitational wave exists. 
By definition, it is characterized by geometric structure 
(metric components and its first derivatives) and 
does not require the premise of asymptotic structure of spacetime. 
So, it is considerably advantageous for the analysis of local spacetime
structure.  

Physical interpretation of the Misner-Sharp energy has been further
backed by various desirable properties: it satisfies the first
law of thermodynamics~\cite{Hayward:1997jp}, 
and it shows the properties of positivity and monotonicity 
under the dominant energy conditions, 
and it reduces to ADM mass in the asymptotically
flat spatial infinity. Reference~\cite{hideki} reinterpreted it 
by the integral of  a locally conserved energy current coming 
from the symplectic structure of spherical symmetry.  
One may find the superiority of the use of Misner-Sharp 
energy in the next section.

The present metric (\ref{metric}) with (\ref{metric_Xi}) gives rise to
\begin{widetext}
\begin{align}
\kappa^2 m = \frac{\pi
 |r|[r^2H_TH_S^5+16t_0^2H_T^2H_S^2-(H_St+3H_Tt_0)^2]}{4t_0^2
 (H_T^{7}H_S^{5})^{1/4}}.
\label{MSenergy}
\end{align}
Physical meaning of each term in this equation is best understood  
at the fiducial time $t=t_0$, at which 
the Misner-Sharp mass is expressed as
\begin{align}
\kappa^2 m(t_0,R) = {\pi(|r|H_S)^3\over 4 t_0^2}
+4\pi{Q\over H_S}+4\pi  Q =\kappa^2\left(
{4\pi\over 3}R^3\,\rho^{(\Phi)}(t_0)
+4\pi \int_Q^R R^2 \D R
\rho^{\rm (em)}(t_0,R)\right)
+4\pi  Q 
\,.
\label{MSenergy_0}
\end{align}
\end{widetext}
The first term corresponds to the energy of the scalar field,
and the second and the last terms to 
the U(1) energies outside of and inside of the black hole,
respectively.

If we set $Q=0$,  the second and third
terms in Eq. (\ref{MSenergy}) are combined to cancel,  and 
$m$ is expressed by the coordinates~(\ref{4D_metric1}) as   
\begin{align}
m=\frac{a r^3}{2 G} \left(\frac{\D a}{\D \bar t}\right)^2
=\frac{4\pi }{3} (ar)^3 \rho^{(\Phi)},
\end{align}
as expected for the background FLRW universe. 
This also justifies the first term in (\ref{MSenergy})
to be the contribution of the scalar field.

%%%%%%%%%%%%%%%%%%%%%%%%%%%%%%%%%%%%%%%%%%%%%%%%%%%%%%%%%%%
\section{Spacetime structure}
\label{sec:spacetime_structure}
%%%%%%%%%%%%%%%%%%%%%%%%%%%%%%%%%%%%%%%%%%%%%%%%%%%%%%%%%%%

Let us now turn to the main task of revealing  
spacetime structure of the solution~(\ref{metric}). 
In order to address this issue, it is of significance to 
discuss the followings: 
\begin{itemize}
 \item Singularity
 \item Trapped surface
 \item Event horizon
 \item Asymptotic structure
\end{itemize}
The first two topics are associated with the
local character of spacetime, 
whereas the last two require global considerations. 
These are elementary issues to be explored in order to characterize 
the spacetime.

We wish to show that our metric~(\ref{metric}) describes a 
black hole in the FLRW universe. To this end, we need to establish 
in the first place that 
the far outside region from the central inhomogeneous domain behaves non-pathologically. 
If the spacetime admits a naked singularity in the asymptotic region--other than the initial big
bang singularity--that are not covered by the event horizon, 
the solution would not gain a popularity as a black hole.

According to a series of theorems due to Penrose and Hawking~\cite{HE},
the appearance of spacetime singularity is closely 
associated to the presence of trapped surfaces.  
Thus, the examination of trapping property may provide us useful 
information of local spacetime geometry.    
In particular, we have two competing effects due to the black hole
and the expanding cosmology: the former tends to 
focus light rays back into the hole while the latter tends to 
spread it out to infinity.
   
At first sight, 
one might expect to gain only a limited perception from the local point of view, 
even though the curvature singularity and local horizons are indeed of
importance. 
Nevertheless, under physically reasonable circumstances, 
the existence of the local horizon implies that the event
horizon lies outside it~\cite{HE}. 
To be more precise, all trapped surfaces are contained
within black holes under the null energy condition 
provided the spacetime is asymptotically flat with some 
additional technical assumptions (see Proposition 9.2.8 in~\cite{HE} for
the proof). 
So, this criterion is of use for our study.

In order to define a black hole as a ``region of no escape,'' 
the spacetime must allow null infinity
as an idealization of an observer sufficiently far in the distance
from some central region. 
Thus the asymptotic analysis should also be put on an emphasis. 
In particular,  we will take a close look at null geodesic
motions, since the null rays play a privileged r\^ole in the 
black hole geometry.
Above listed considerations are sufficient to provide us
with insight into the global pictures of our dynamical spacetime~(\ref{4D_metric1}).

In the remainder of this paper, we shall simplify our notations 
by using dimensionless variables 
$\tilde t:={t/t_0}$ and  $\tilde r := {r/ Q}$. 
The metric is also rewritten into dimensionsless form
$\D \tilde s_4^2=Q^{-2}\,\D s_4^2 $
as
%------------- Dimensionless metric  ----------------%
\begin{eqnarray}
\D \tilde s_4^2 =-\tau^2 \, \tilde\Xi \, \D \tilde t^2
+\tilde \Xi^{-1}\left(\D \tilde r^2+\tilde r^2 \D \Omega_2^2\right)
\,,
\label{4D_metric0}
\end{eqnarray}
with
\begin{eqnarray}
\tau :={t_0\over Q}
\,,\qquad 
\tilde\Xi :={\tilde r^2\over [ (1+\tilde t\tilde r)(1+\tilde r)^3]^{1/2}}
\,.
\label{tau_Xi}
\end{eqnarray}
In this form, the metric
involves only one dimensionless parameter $\tau=t_0/Q$, 
the physical meaning of
which will be revealed below.
We will affix ``tilde'' to denote dimensionless quantities in what follows.

%%%%%%%%%%%%%%%%%%%%%%%%%%%%%%%%%%%%%%%%%%%%
\subsection{Singularity}
%%%%%%%%%%%%%%%%%%%%%%%%%%%%%%%%%%%%%%%%%%%%
\label{sec:singularity}

The scalar curvature and the Kretschmann invariant scalar
are given by
\begin{widetext}
\begin{align}
{\cal R}=&
\kappa^2  \left(P_\theta^{(\Phi)}+P_\phi^{(\Phi)}\right)
={3\over 8Q^2}\left[
{1\over \tau^2}\left({H_T\over H_S}\right)^{3/2}-
{1\over \tilde r^4}{(H_T-H_S)^2\over (H_T^{5}H_S^{7})^{1/2}}
\right], 
\\
{\cal R}_{\mu\nu\rho\sigma}{\cal R}^{\mu\nu\rho\sigma}
=& {1\over 64 \tau^4 Q^4 H_T^5 H_S^7  }
\Biggl[15H_T^2H_S^{10}
+6\left({\tau^2 \over \tilde r^4}\right)H_T H_S^5
\left(7H_T^2+10 H_T H_S-H_S^2\right)
%\right. 
\nonumber \\
&+
\left({\tau^4 \over \tilde r^6}\right)\left\{
9(31H_S^2+2H_S+159)H_T^4
+12(7H_S^2+
9
H_S+
96)H_T^3H_S
\right.
\nonumber \\
&+
%\left.
\left.
6(15H_S^2-126H_S+143)H_T^2H_S^2
-12(H_S-1)(H_S+15)H_TH_S^3
+71(H_S-1)^2H_S^4
\right\}
\Biggr]\,.
\label{Kretschmann}
\end{align}
\end{widetext}
These curvature invariants diverge when $H_T=0$ and $H_S=0$, that is,
at 
\begin{eqnarray}
 \tilde t=\tilde t_s(\tilde r ):=-1/\tilde r, \qquad {\rm and }
\qquad  \tilde r=-1 \,.
\end{eqnarray}
At these spacetime points, the circumference radius $R$, Eq.~(\ref{R}), vanishes, i.e.,
they are central shell-focusing singularities. 
Thus, around infinity is free from singularities and is well-behaved. 

\begin{figure}[h]
\begin{center}
\includegraphics[width=6.5cm]{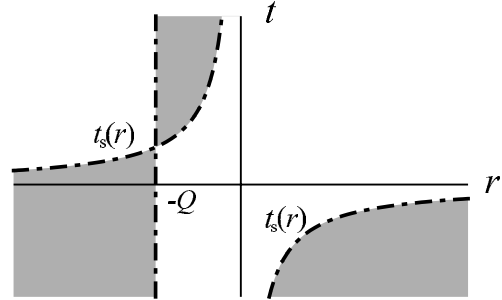}
\caption{
Allowed coordinate ranges. 
The grey zone denotes the forbidden region, and the dashed curves 
correspond to curvature singularities.}
\label{fig:allowed}
\end{center}
\end{figure}

It deserves to observe that 
$\tilde t=0$ surface, 
where the scale factor $a(\bar t)$ appearing in the 
metric (\ref{4D_metric1}) vanishes, is not singular at all 
since the curvature invariants remain finite therein. 
It follows that the big bang singularity $\tilde t=0 $ is smoothed out 
due to a nonvanishing Maxwell charge  $Q~(>0)$.  
Hence, one has also  to consider the $\tilde t<0$ region  in the coordinates (\ref{metric}). 
In addition, we find that the $\tilde r=0$ surface is neither singular, 
thereby we may extend the spacetime across the $\tilde r=0$ surface to
$\tilde r<0$.
Since the allowed region is where  $H_TH_S^3 >0$ is satisfied, 
we shall focus attention to the coordinate domain
\begin{align}
\tilde t\ge \tilde t_s(\tilde r),\qquad
\tilde r\ge -1\,, 
\end{align}
in the subsequent analysis.
Another permitted region $\tilde t>\tilde t_s$ and 
$\tilde r<-1$ is not our immediate interest here, 
since it turns out to be causally disconnected to the outside region, 
as we shall show below. 
Possible allowed coordinate ranges are depicted in 
Figure~\ref{fig:allowed}.

%-------------------------------------------%

Since our spacetime is spherically symmetric, 
electromagnetic and gravitational fields do not radiate. 
Thereby, it is more advantageous to concentrate on   
their ``Coulomb components.''  
For this purpose, let us introduce the 
Newman-Penrose  null tetrads by 
\begin{align}
&
l_\mu\D \tilde x^\mu=\sqrt{\frac{\Xi}{2}}(-\tau 
\D \tilde t+\Xi^{-1}\D \tilde r), 
\nonumber \\
&
 n_\mu\D \tilde  x^\mu=
\sqrt{\frac{\Xi}{2}}
(-\tau \D \tilde t-\Xi^{-1}\D \tilde r),
\label{NPtetrads} \\
&
 m_\mu\D \tilde  x^\mu =\frac {\tilde  r}{\sqrt{2\Xi}}
(\D \theta + i \sin \theta \D \phi ).
\nonumber
\end{align}
with $\bar m_\mu$ being a complex conjugate of $m_\mu$.
They satisfy the orthogonality conditions 
$l^\mu n_\mu=-1=-m^\mu\bar m_\mu$ and $l^\mu l_\mu=n^\mu n_\mu =m^\mu m_\mu=\bar m^\mu\bar m_\mu=0$.
Since $\tilde t$ is a timelike coordinate everywhere, $l^\mu $ and $n^\mu$
are both future-directed null vector orthogonal to metric spheres.

The only nonvanishing Maxwell and Weyl scalar are their 
``Coulomb part,''
$\phi_1^{(A)}:=-\frac 12F_{\mu\nu}^{(A)}(l^\mu n^\nu +\bar m^\mu m^\nu)$
and 
$\Psi_2 :=-C_{\mu\nu\rho\sigma}l^\mu m^\nu \bar m^\rho n^\sigma$, 
both of which are invariant under the tetrad transformations
due to the type D character.
It is readily found that
\begin{align}
 \phi^{(T)}_1 =\frac{\sqrt \pi }{\sqrt 2 \kappa Q\tilde r^2 H_T^2}, \qquad 
  \phi^{(S)}_1 =\frac{\sqrt \pi }{\sqrt 2 \kappa Q\tilde r^2 H_S^2}\,,
\end{align}  
and 
\begin{align}
 \Psi_2
 =\frac{\Xi_{,\tilde r}-\tilde  r\Xi_{,\tilde  r\tilde r}}{6Q^2 \tilde r}
=\frac{6 \tilde r H_T^2+(H_T-H_S)^2+2\tilde t  \tilde r H_S^2
}{8Q^2 \tilde r^4(H_T^{5}H_S^{7})^{1/2}}\,.
\end{align}
The loci of singularities at which these quantities diverge are 
the same as the positions of the above singularities.
One may also recognize that at the fiducial time $\tilde t=1$, 
above curvature invariants are the same as the
extremal RN solution, as expected.

Let us next look into the causal structure of singularities. 
Since the Misner-Sharp mass~(\ref{MSenergy}) becomes negative
as approaching these singularities [the third term in Eq.~(\ref{MSenergy})
begins to give a dominant contribution], 
we speculate from our rule of thumb that these singularities are
both contained in the untrapped region and possess the timelike structure.

To see this more rigorously, 
let us consider radial null geodesics 
%--------------- radial null  ----------------%
\begin{align}
\left. \frac{\D \tilde r}{\D  \tilde t}\right|_\pm 
=\pm \tau  \tilde  \Xi \,,
\label{radial_null}
\end{align}
in the neighborhood of these singularities. 
Here, the upper (lower) sign corresponds to the geodesics along the 
direction $l^\mu$ ($n^\nu$), which
we shall refer to as outgoing (ingoing).  
If we can find an infinite number of null geodesics that emanate from and terminate into 
the singularity, the singularity turns out to be timelike. 
If there exists a unique geodesic terminating into (emanating from) the
singularity and an infinite number of geodesics emanating from
(terminating into) it, the singularity has an ingoing (outgoing) null
structure. Whereas, if there exist an infinite number of 
ingoing and outgoing geodesics terminating into or emanating from the 
singularity, we can conclude that it is spacelike.

We begin by the analysis of the singularity at $ \tilde r=-1$. 
In the vicinity of the singularity $ \tilde r=-1$, 
we suppose that the null geodesics have the following asymptotic
solution, 
\begin{align}
 \tilde r+1 = C_1 \left|
 \tilde t-\tilde t_1
%\frac{t-t_1}{t_0}
\right|^p,
\label{asy_geodesics1}
\end{align}
where $C_1 $, $\tilde t_1$ and $p$ are constants. 
$C_1$ is taken to be positive since we are concerned with the
region $ \tilde r>-1$. $\tilde t_1$ can be regarded as the arrival
time of the ingoing null geodesics at the singularity, 
or the departure time of the outgoing null geodesics from the singularity. 
Consider first the $ \tilde t>\tilde t_1$ case, that is, the 
geodesics emanating from the singularity.  
Substituting the assumed form~(\ref{asy_geodesics1})
into~Eq. (\ref{radial_null}), 
we find that only  the outgoing 
null geodesics  have the solution, for which
\begin{align}
p=\frac{2}{5}, \qquad
C_1^{5/2} =
\frac 52\tau (1-\tilde t_1)^{-1/2}
\,.
\label{Cpt1}
\end{align} 
This reveals that for $ \tilde t>\tilde t_1$ there exist radial null
geodesics that departed the singularity at $ \tilde t=\tilde t_1$. 
Following the identical procedure, 
it can be shown that for $ \tilde t<\tilde t_1$ 
only the ingoing null geodesics have the solution for which 
$C_1$ and $p$ are given by~(\ref{Cpt1}). 
This means that there exist radial null
geodesics that will get to the singularity at $ \tilde t=\tilde t_1$.
Therefore, we establish that the singularity $ \tilde r=-1$ is truly timelike, 
that is, a locally naked singularity, 
since we have only one set of outgoing and ingoing null solutions
parametrized by their arrival or emanating time.   
It is also obvious from Eq.~(\ref{Cpt1}) that the singularity 
$ \tilde  r=-1$ exists only for $ \tilde t<1$.

In an analogous fashion, assume 
the asymptotic form of the null geodesics
near the singularity $ \tilde r=\tilde r_s (=-1/ \tilde t)$ as
\begin{align}
 \tilde r+\frac{1}{ \tilde  t}
%\frac{t_0Q}{t}
=C_2\left| \tilde t-\tilde t_2\right|^q\,,
\label{asy_geodesics2} 
\end{align}
where $C_2$, $\tilde t_2$ and $q$ are constants. 
In the present case, $C_2$ takes positive (negative) value  
for $\tilde t_2 > 1 ~(\tilde t_2< 1) $. 
\begin{widetext}
Plugging this into Eq.~(\ref{radial_null}), we obtain $q=2/3$ and 
\begin{subequations}
\label{Cqt2}
\begin{align}
C _2 ^{3/2} &= \frac{3}{2}\tau  
\left[\tilde t_2^2 
(\tilde t_2-1)^3\right]^{-1/2},
\qquad \textrm{for outgoing null with $\tilde t>\tilde t_2$}\,,\label{Cqt2a}\\
 (-C _2) ^{3/2}&= \frac{3}{2}\tau \left[\tilde t_2^2 
(1-\tilde t_2)^3\right]^{-1/2},
\qquad \textrm{for
ingoing null with $\tilde t>\tilde t_2$}\,, \label{Cqt2b}\\
C _2 ^{3/2}&= \frac{3}{2}\tau 
\left[\tilde t_2^2 
(\tilde t_2-1)^3\right]^{-1/2},
 \qquad \textrm{for ingoing null with $\tilde t<\tilde t_2$}\,, \label{Cqt2c}\\
 (-C _2) ^{3/2}&= \frac{3}{2}\tau
 \left[\tilde t_2^2 
(1-\tilde t_2)^3\right]^{-1/2},
\qquad \textrm{for
outgoing null with $\tilde t<\tilde t_2$}\,. \label{Cqt2d}
\end{align}
\end{subequations}
\end{widetext}
Eqs.~(\ref{Cqt2b}) and (\ref{Cqt2d}) indicate that 
there exist null geodesics that originate from and 
come to the singularity at any time $\tilde t_2<0$. 
Thus, the singularity $\tilde t=-1/\tilde r ~(<0)$ appearing in the 
$ \tilde r>0$ region is also timelike. 
Whereas, 
Eqs.~(\ref{Cqt2a}) and (\ref{Cqt2c}) show that 
the timelike singularity $ \tilde r=-1/ \tilde t ~(<0)$
occurs only for  $ \tilde t >1$.

Although 
we have not dwelt on the 
existence proof of the asymptotic solutions
[Eqs~(\ref{asy_geodesics1}) with (\ref{Cpt1}) and
(\ref{asy_geodesics2}) with (\ref{Cqt2})] of the radial null geodesics, 
the proof 
can be obtained via the contraction mapping method following 
the argument in 
 e.g., the appendix of~\cite{Nozawa}. It is also noted that 
it is sufficient to focus attention on the radial null geodesics in the
vicinity of central singularities: causal geodesics excluding radial
null geodesics will
fail to emanate from singularity if radial null geodesics 
do not arise (see also the appendix in~\cite{Nozawa}).

%%%%%%%%%%%%%%%%%%%%%%%%%%%%%%%%%%%%%%%%%%%%%%%%%%%%%
%%%%%%%%%%%%%%%%%%%%%%%%%%%%%%%%%%%%%%%%%%%%%%%%%%%%%
\subsection{Trapping horizons}
%%%%%%%%%%%%%%%%%%%%%%%%%%%%%%%%%%%%%%%%%%%%%%%%%%%%%
%%%%%%%%%%%%%%%%%%%%%%%%%%%%%%%%%%%%%%%%%%%%%%%%%%%%%
\label{sec:TH}

Our primary concern in this article is to reveal that the 
spacetime describes a black hole. However, this is not only technically
 but also conceptually difficult.   
Since a black hole is by definition a ``region of no escape,''
the locus of event horizon as its boundary has a teleological meaning. 
Unless we know the entire future of our universe, 
we are unable to determine whether a ``black hole candidate''
is qualified as a black hole. The concept of the black hole is 
considerably messy in practical point of view.

In physically acceptable situations, however, we can rely on the 
the notion of {\it trapped surface}, first introduced by Penrose~\cite{Penrose:1964wq}.
The concept of trapped surface is inherently local, 
such a difficulty does not arise.  
Imagine a massive star undergoing a gravitational collapse  to form a black
hole. There appears a region for which even ``outgoing'' null rays are
dragged  back due to strong gravity  and have a negative expansion. For each time slice, 
this defines an {\it apparent horizon}~\cite{HE} as an outermost boundary of the trapped region 
in the asymptotically flat spacetimes. 
Hayward generalized these quasilocal concepts to define a 
class of {\it trapping horizons}~\cite{Hayward1993}.  
One strength of the use of trapping horizons is just to 
encompass various types of horizons associated not only with black holes 
but also with white holes and cosmological ones.   
As commented at the beginning of this section, the underlying aim of
this direction is to gain useful guide for event horizon from these
local analysis.

Let us consider a compact spacelike orientable surface $S$. We take $S$
as a metric sphere, respecting an SO(3)-symmetry of background
universe. It then follows that 
the Newman-Penrose tetrads $(l^\mu, n^\mu)$ 
defined in Eq.~(\ref{NPtetrads})
are normal to $S$ (i.e., they are radial) and future-directed null vectors. 
Due to the spherical symmetry, they are shear-free and rotation-free. 
Define the associated 
null expansions $\theta_\pm $ by
%-------------- Expansions ------------%
\begin{align}
\theta _+ :=2m^{(\mu}\bar m^{\nu)}\nabla_\mu l_\nu, \qquad
 \theta _- :=2m^{(\mu}\bar m^{\nu)}\nabla_\mu n_\nu. 
\end{align} 
In the coordinates~(\ref{4D_metric0}), they are expressed as
%-------------- Expansions ------------%
\begin{align}
\tilde \theta_\pm =\frac{\tilde rH_S(H_TH_S^3)^{1/2}
 \pm \tau(3 H_T+H_S\tilde t)}
{2\sqrt 2 \tau \tilde r (H_T^{5}H_S^{7})^{1/4}}.
\label{expansions}
\end{align}
Note that the signs of $\theta_\pm $ have an invariant meaning, but
each value of $\theta _\pm$ is not a universal quantity due to the 
``class III'' tetrad rotations. An invariant combination is their product 
$\theta_+\theta_-=-2R^{-2}(\nabla_\mu R)(\nabla^\mu R)$.

Expansions $\theta_\pm$ characterize the extent to which the 
light rays are diverging or converging, or equivalently the rate at
which the area of metric sphere is increasing or decreasing in the null
directions.
In terms of null expansions $\theta_\pm$,    
a metric sphere is said to be {\it trapped} ({\it untrapped}) 
if $\theta_+\theta_- > 0$ ($\theta_+\theta_- < 0$), 
and {\it marginal} when $\theta _+\theta_- =0$.  
A marginal surface is said to be 
{\it future} ({\it past}) if $\theta _+=0~ (\theta_-=0)$. 
A future marginal surface is further classified into {\it outer } ({\it inner})
if $n^\mu \nabla_\mu \theta _+<  0$
($n^\mu \nabla_\mu \theta _+> 0$)
 and similarly a past marginal surface is
called {\it outer } ({\it inner}) if 
$ l^\mu \nabla_\mu  \theta _-> 0$
($ l^\mu \nabla_\mu  \theta _-<0$).
A {\it trapping horizon} is the closure of a hypersurface foliated by future
or past, and outer or inner marginal surfaces~\cite{Hayward1993}. 
In terms of the Misner-Sharp energy, $R< 2Gm$ 
($R>2Gm$) defines the trapped (untrapped)
region and  
the marginal surface ($\theta _+\theta_- =0$) is positioned at 
the ``Schwarzschild radius'' $R=2Gm$.  
A ``normal'' spacetime region as occurred in the flat space 
is composed of untrapped surfaces on which 
outgoing rays have positive expansions while 
the ingoing rays have negative expansions.

Among these classes of trapping horizons, the 
future-outer trapping horizons turn out to be most relevant 
in the context of black holes. 
The future outer trapping horizon properly captures the intuitive idea that the ingoing null rays are
converging with the outgoing null ray being instantaneously  parallel on
the horizon, diverging outside and converging inside.  
Inner trapping horizons are associated with 
cosmological horizons, and interior horizons of black holes. 
The past trapping horizons arise  when discussing white holes and 
cosmological ones. 
Since the concept of trapping horizons is sufficiently general,
it is considerably useful for the analysis of black holes especially in the 
non-asymptotically flat  spacetimes.

The expansions are intimately associated to the 
variation of the Misner-Sharp energy through the 
first law~\cite{Hayward1994}, 
\begin{align}
 l^\mu \nabla_\mu m&=2\pi R^3(T_{\mu \nu }l^\mu n^\nu 
\theta_ + -T_{\mu \nu }l^\mu l^\nu \theta_- ),
\nonumber \\
 n^\mu \nabla_\mu m&=2\pi R^3(T_{\mu \nu }l^\mu n^\nu 
\theta_ - -T_{\mu \nu }n^\mu n^\nu \theta_+ )\,.
\label{variationFormula}
\end{align}
The present system satisfies the dominant energy condition, 
which implies $T_{\mu \nu }l^\mu l^\nu \ge 0$, 
$T_{\mu \nu }n^\mu n^\nu \ge 0$ and $T_{\mu \nu }l^\mu n^\nu \ge 0$.
Hence, Eq.~(\ref{variationFormula}) establishes that
$m$ is not decreasing (not increasing) along the 
$l^\mu $-direction ($n^\mu $-direction) in the untrapped region
of $\theta_+>0$ and $\theta_-<0$. 
This illustrates that the Misner-Sharp energy is a monotonically
increasing function toward outwards  
in the ordinary region where $\theta_+>0$ and $\theta_-<0$ are satisfied, 
which accords with our intuition.

Let us now investigate the properties of 
trapping horizons more closely.
From Eq.~(\ref{expansions}),  
trapping horizons occur at 
\begin{align}
\tilde rH_S(H_TH_S^{3})^{1/2}\pm \tau(3 H_T+H_S\tilde t)=0,
\label{THeq}
\end{align}
where the upper (lower) sign corresponds to 
$\tilde \theta_+=0$
($\tilde \theta_-=0$). Noticing that $\tilde rH_S\ge 0$, 
Eq.~(\ref{THeq}) is solved as 
$\tilde t=\tilde t^{(\mp)}_{\rm TH}(\tilde r)$,
 where
%-------------- Trapping horizons --------------%
\begin{widetext} 
\begin{align}
\label{TH}
\tilde t_{\rm TH}^{(\mp)}(\tilde r):=&
\frac{\tilde r^2}{2\tau^2(H_S+3)^2}
\left[H_S^5-6\tau^2(H_S+3)\tilde r^{-3}
\mp H_S^3  
\sqrt{H_S^4+4\tau^2(H_S+3)\tilde r^{-3}}\right].
\end{align}
\end{widetext} 
Here, $\tilde \theta_\pm=0$ holds at 
$\tilde t=\tilde t^{(\mp)}_{\rm TH}(\tilde r)$.
In the  pursuing subsection, 
we shall separately
analyze   the $\tilde r> 0$ and $\tilde r<0$ cases corresponding to
outside and inside the black-hole event horizon.

\begin{figure}[t]
\begin{tabular}{cc}
\includegraphics[width=6cm]{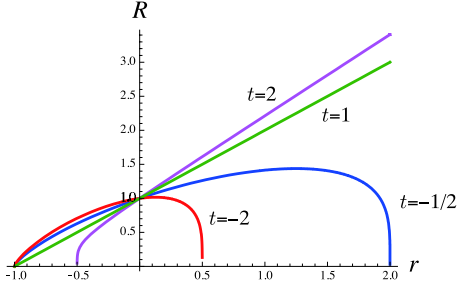}
\end{tabular}
\caption{
Typical behaviors of 
$\tilde R=[(\tilde t\tilde r+1)(\tilde r+1)^3]^{1/4}$
as a function of $\tilde r$ with fixed time. 
For fixed $\tilde t<0$, the circumference radius $\tilde R$ ceases to
 increase monotonically  with $\tilde r$, 
but has a maximum at $\tilde t=\tilde t_c(\tilde r)$. 
For $\tilde t>0$, $\tilde R$ turns to be monotonic in $\tilde r$.
}
\label{fig:Sr}
\end{figure}

Before going into the detailed argument,
we pause for a moment to discuss the behavior of $\tilde R(\tilde t, \tilde r)$. 
We have taken $l^\mu$ (and correspondingly $n^\mu$) in such a way that 
$\tilde r$ increases (decreases) along $l^\mu ~(n^\mu )$. 
From 
%---------------- R,t and R,r ---------------%
\begin{align}
\tilde R_{,\tilde t} =\frac{|\tilde r| }{4}
\left(\frac{H_S^3}{H_T^3}\right)^{1/4},\quad
\tilde R_{,\tilde r} =\frac{|\tilde r|[3H_T+H_S\tilde t]}
{4\tilde r(H_T^{3}H_S)^{1/4}}, 
\end{align}
one finds that 
$\tilde R_{,\tilde r}> 0$
($\tilde R_{,\tilde r}< 0$) holds for 
$\tilde t > \tilde t_{c}(\tilde r)$
($\tilde t < \tilde t_{c}(\tilde r)$), where
%------------- critical surface --------------%
\begin{align}
\tilde t_{c}(\tilde r):=-\frac{3}{4\tilde  r+1 }.
\end{align}
This means that $\tilde R $ increases as $\tilde r$ grows for 
$\tilde t>\tilde t_c$, deserving 
$l^\mu $ to be called ``outgoing.'' 
However,  this is no longer true for $\tilde t<\tilde  t_c (\tilde r)$. 
This means that there exists a maximum value of 
$\tilde R(\tilde  t, \tilde r)$ for given time (see Figure~\ref{fig:Sr}).
It should be also remarked that the invariant scalar curvatures are  all
finite at this surface $\tilde t=\tilde t_c(\tilde r)$, this is not the
shell-crossing singularity.

\subsubsection{Trapping horizons in the region of $\tilde r> 0$}

\begin{center}
\begin{figure}[t]
\includegraphics[width=5.5cm]{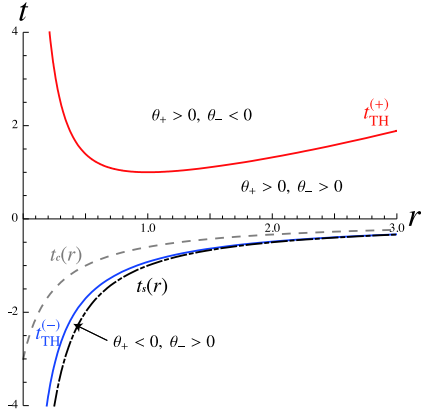}
\caption{Typical curves of trapping horizons 
$\tilde t^{(\pm)}_{\rm TH}(\tilde r)$
for $\tau <\tau _{\rm crit}$ outside the event horizon $\tilde r>0$.
The plot is $\tau =1$.  
$\tilde t^{(+)}_{\rm TH}$ first occurs at some time (in this
 case $\tilde t=\tilde r=1$) and bifurcates into two branches.  
The black dashed line and the grey dotted line denote $\tilde t_s(\tilde r)$ and 
$\tilde t_{c}(\tilde r)$, respectively.}
\label{fig:TH}
\end{figure}
\end{center}

Let us begin by the case of  $\tilde r> 0$.
After simple calculations, one obtains
\begin{widetext} 
\begin{align}
\tilde  t_{\rm TH}^{(+)}-\tilde t_{c}(\tilde r)&= 
\frac{\tilde r^2}{2\tau 
(H_S+3)^2}\left(H_S^3\sqrt{H_S^4+{4\tau^2}{\tilde r^{-3}}(H_S+3)}
+H_S^5\right)> 0\,, \\
\tilde  t_{c}(\tilde r)-\tilde t_{\rm TH}^{(-)}&= 
\frac{r^2}{2\tau (H_S+3)^2}\left(H_S^3\sqrt{H_S^4+{4\tau^2}{\tilde 
r^{-3}}(H_S+3)}
-H_S^5\right)> 0\,, 
\label{THminusTc}\\
\tilde t_{\rm TH}^{(-)}-\tilde t_s (\tilde r)&=
\frac{\tilde r^2}{2\tau (H_S+3)^2}
\left[
\frac 12H_S\left(H_S^2-\sqrt{H_S^4+4\tau^2\tilde r^{-3}(H_S+3)}\right)^2
+\frac 12 H_S^5+\frac{\tau^2}{\tilde r^2}H_S(H_S+3)
\right]> 0\,,
\end{align}
\end{widetext} 
for $\tilde r> 0$.
Figure~\ref{fig:TH} shows the typical curves of trapping horizons 
$\tilde t^{(\pm )}_{\rm TH}$. 
The region $\tilde t_{\rm TH}^{(-)}<\tilde t<\tilde t^{(+)}_{\rm TH}$ denotes
a past trapped region in which even ingoing light rays are
diverging due to the cosmic expansion.

Next, let us delve deeper into the type of trapping horizon.
It is straightforward to find that 
\begin{align}
l^\mu\tilde \nabla_\mu \tilde \theta_- =\left.\frac{F}{4\tilde  r^4(H_T^5H_S^{7})^{1/2}}
\right|_{\tilde t=\tilde t_{\rm TH}^{(+)}},
\label{THtype1}
\end{align}
along the trapping horizon with $\tilde \theta_-=0$.
Here, we have introduced a function
\begin{align}
F:= H_S^2-(4\tilde r^2+8
\tilde  r+1)H_T^2\,, 
\end{align}
the sign of which controls the type of trapping horizon. 
For $\tilde r>0$,  the inequality
$\tilde r+1<\sqrt{4\tilde  r^2 +8 \tilde  r+1}$
holds,  from which we obtain 
\begin{align}
F
<- \tilde t (H_T+1/r)(4\tilde r^2+8 \tilde r+1)<0\,.
\end{align}
Hence, for the trapping horizon with $\tilde t^{(+)}_{\rm TH}>0$, 
$l^\mu \tilde \nabla_\mu \tilde \theta_- <0$ is satisfied.   
That is to say, the past trapping horizon $\tilde t^{(+)}_{\rm TH}$ 
occurred in the $\tilde r>0, \tilde t>0$
region is always of outer-type. This may be ascribed to the 
cosmological origin  
since a past-outer trapping horizon develops when
the background FLRW universe is filled with a
stiff matter.  More stringent bound for the 
condition $F>0$ is numerically found to be 
$\tau < \tau_{\rm crit} \sim 5.444$.
In the case of $\tau> \tau_{\rm crit}$, 
a part of the past trapping horizon near $\tilde r=0$
becomes inner rather than outer.

Similarly, we obtain 
\begin{align}
 n^\mu\tilde \nabla_\mu \tilde \theta_+
=\left.\frac{F}{4 \tilde  
r^4(H_T^5H_S^{7})^{1/2}}
\right|_{\tilde t=\tilde t_{\rm TH}^{(-)}}\,,
\label{THtype2}
\end{align}
along the trapping horizon with $\tilde \theta_+=0$. 
From Eq. (\ref{THminusTc}), we find 
\begin{align}
F|_{\tilde t=\tilde t^{(-)}_{\rm TH}(\tilde r)}
&>\left.H_S^2-(4\tilde r^2+8 \tilde r+1) 
H_T^2\right|_{\tilde t=\tilde t_c}\nonumber \\
&=12 H_S^2(3+H_S)^{-2}>0\,.
\end{align}
Hence, 
the trapping horizon $\tilde t^{(-)}_{\rm TH}$
is necessarily of future-inner type. 
The appearance of untrapped region $\tilde \theta_+<0 $ 
and $\tilde \theta_->0$ is due to the repulsive nature of the 
timelike naked singularity.

From the general argument, the outer (inner) trapping horizons 
must be non-timelike (non-spacelike) under the null energy 
condition~\cite{Hayward1994,Nozawa}. So the above analysis asserts
that the trapping horizon $\tilde t^{(+)}_{\rm TH}$ 
never becomes timelike for $\tau < \tau_{\rm crit} $, while 
$\tilde t^{(-)}_{\rm TH}$ cannot be spacelike anywhere.

Here, it is worthwhile to remark the behavior
of trapping horizons in various asymptotic limits. 
In the $\tilde r\to 0$ limit, 
Eq.~(\ref{TH}) yields that 
$\tilde r \tilde t^{(\pm)}_{\rm TH}$ 
is finite  and given by
%------------- r-> 0 limit of TH  --------------%
\begin{align}
&\tilde r \tilde t^{(+)}_{\rm TH} = \frac{1+ \sqrt{1+4\tau ^2}}{
2\tau^2 }> 0\,, 
\label{r0limit1}
\\
&
\tilde r \tilde t^{(-)}_{\rm TH} = \frac{1- \sqrt{1+4\tau ^2}}{
2\tau^2 }< 0\,,
\label{r0limit2}
\end{align}
as $\tilde r\to 0$.
Using this equation, one can find that  the circumference radii of the 
trapping horizons $\tilde t^{(\pm)}_{\rm TH}$ respectively 
approach to some constants $\tilde R_\pm $
as $\tilde r\to 0$, where
%-------------  R_pm  ---------------------%
\begin{align}
 \tilde R_\pm :=  \left(\frac{\sqrt{1+4\tau^2 }\pm 1}{2\tau
 }\right)^{1/2}\,. 
\label{R_plusminus}
\end{align}
These surfaces correspond to the infinite redshift $(\tilde t\to \infty) $ and blueshift 
$(\tilde t\to -\infty )$  surfaces  with respect to an asymptotic
observer. We will see in the next section that these surfaces represent the
black hole and white hole horizons. 
Notice that $\tilde R_+ $ (and respectively $\tilde R_-$) is 
a monotonically decreasing (increasing)
function of $\tau $ and they behave as 
$\tilde R_+ \to \infty $ and $\tilde R_-\to 0$ 
in the limit $\tau\to  0$, while they 
asymptotically tend to unity as $\tau \to \infty $. 
According to Eq. (\ref{R_plusminus}), 
$Q$ and $\tau$ are expressed in terms of $R_\pm$ as 
%------------- Q and tau  ----------------%
\begin{align}
Q=\sqrt{R_+R_-}, \qquad
\tau =\frac{R_+R_-}{R_+^2-R_-^2}\,. 
\label{Qtau}
\end{align}
Hence, we find that the charge $Q$ sets the geometrical mean of
horizon radii and  
their relative ratio is encoded in the 
parameter $\tau $.

The physical meaning of $\tau$ is found by evaluating
the energy densities of the dilaton field and U(1) field
at the horizon $R_+$ as follows:
Those densities are given by
%------------ energy densities at H  --------------%
\begin{align}
&\kappa^2\rho^{(\Phi)}|_{R_+}
={3\over 8t_0^2\tilde R_+^6}\,,
\\
&\kappa^2\rho^{\rm (em)}|_{R_+}
={1+3\tilde R_+^{8}\over 4Q^2\tilde R_+^{10}}
\,.
\end{align}
Then, the ratio is found to be
\begin{align}
{\rho^{\rm (em)}
\over \rho^{(\Phi)}}\Big{|}_{R_+}
={2\tau^2 (1+3\tilde R_+^{8})\over 3\tilde R_+^4}
\,.
\end{align}
From the expression of $R_+$, we find that $\tau $ has a 
one-to-one correspondence to  $\rho ^{(\rm em)}/\rho^{(\Phi)}|_{R_+}$
and is given by
\begin{align}
\tau^2&={1\over 8}\left[3\left({\rho^{\rm (em)}
\over \rho^{(\Phi)}}\Big{|}_{R_+}-1\right)-\sqrt{3\left(2\,{\rho^{\rm (em)}
\over \rho^{(\Phi)}}\Big{|}_{R_+}-1\right)}
\right]\,,
\nonumber 
\\
&\approx
\left\{
\begin{array}{ccc}
{3\over 8}
\,{\rho^{\rm (em)}
\over \rho^{(\Phi)}}\Big{|}_{R_+}
 & {\rm for} &
{\rho^{\rm (em)}
\over \rho^{(\Phi)}}\Big{|}_{R_+}
\gg 2\,,
\\
{1\over 4}\left({\rho^{\rm (em)}
\over \rho^{(\Phi)}}\Big{|}_{R_+}-2\right) & {\rm for} &
{\rho^{\rm (em)}
\over \rho^{(\Phi)}}\Big{|}_{R_+}
\sim 2\,.
\end{array}
\right.
\nonumber
\\~
\end{align}
$\tau$ is a monotonic function of the ratio, and
it vanishes when the ratio approaches $2$. 
The ratio $(\rho^{\rm (em)}/\rho^{(\Phi)})|_{R_+}$,
which must be larger than 2 for $\tau^2>0$,
corresponds to $\tau$ by a one-to-one mapping.
Hence $\tau$ is related to the ratio of 
two densities at the horizon.

In the limit $\tilde r\to 0$, one finds that  
$\tilde \theta_+  $ and $\tilde \theta _-$ both 
vanish, implying that the  $\tilde r=0$ surface becomes degenerate into an
ingoing and outgoing null structure.

On the other hand,
taking $\tilde r\to \infty $, $\tilde t^{(+)}_{\rm TH}$ diverges as
$(\tilde r^2/4\tau)^2$, whence
$\tilde R (\tilde t^{(+)}_{\rm TH} (\tilde r), \tilde r) 
\to \tilde r^3/4\tau \to \infty $. 
While we have
$\tilde t^{(-)}_{\rm TH}\to -1/\tilde r= \tilde t_{s}(\tilde r)$
  as $\tilde r\to \infty $. 
But this does not mean that the trapping horizon 
$\tilde t^{(-)}_{\rm TH}$ plunges into the singularity $\tilde t_s$ 
as $\tilde r\to \infty$. One can verify by taking higher order terms into
account that $\tilde t^{(-)}_{\rm TH}$ tends to have a constant radius
$\tilde  R (\tilde t^{(-)}_{\rm TH} (\tilde r), \tilde r)\to \sqrt{\tau}$ in this limit.

\subsubsection{Trapping horizons in  the region of $r<0$}

\begin{figure}[t]
\bigskip
\begin{center}
\includegraphics[width=5.5cm]{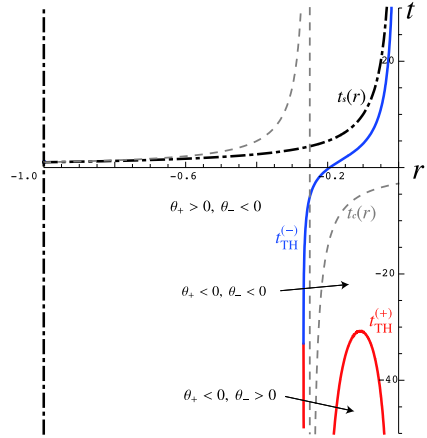}
\caption{Typical curves of trapping horizons 
$\tilde t^{(\pm)}_{\rm TH}(\tilde r)$
for $\tau<\tau_{\rm crit}$ inside the event horizon $\tilde r<0$. 
The plot is for $\tau=1$, for which the trapping horizon 
$\tilde t^{(+)}_{\rm TH}$ has a maximum
 at $\tilde r\simeq -0.1080$,  spacelike for 
$-0.1299\lesssim \tilde r<0$ and merge with $\tilde t^{(-)}_{\rm TH}$
at $\tilde r_0\simeq -0.2679$. 
}
\label{TH_negativer}
\end{center}
\end{figure}

For negative values of $\tilde r$, 
$\tilde t^{(-)}_{\rm TH}>\tilde t^{(+)}_{\rm TH}$ holds,
in contrast to the $\tilde r>0$ case. 
Two trapping horizons $\tilde t^{(\pm)}_{\rm TH}$ develop  
only in the region 
$\tilde r_0<\tilde r< 0 $ and they
coincide at the point $\tilde r=\tilde r_0$, where the square-root of Eq.~(\ref{TH})
vanishes. 
Namely, $\tilde r_0$ satisfies 
\begin{align}
I(\tilde r_0):= H_S^4(\tilde r_0)+4\tau^2(H_S(\tilde r_0)+3)\tilde r_0^{-3}=0\,. 
\end{align} 
Since $I(-1)=-12\tau^2 <0$ and 
$I(-1/4)=81 >0$, $-1<\tilde r_0<-1/4$ is concluded. 
We can also find that 
the trapping horizon $\tilde t^{(+)}_{\rm TH}$ negatively diverges 
at $\tilde r=-1/4$. 
Figure~\ref{TH_negativer} plots typical curves of trapping horizons
occurred in $\tilde r<0$.

Equations (\ref{THtype1}) and (\ref{THtype2}) continue to be true for
$\tilde r<0$. We can find numerically that
at  $\tilde t^{(+)}_{\rm TH}(\tilde r)$ $F<0$ holds around 
$\tilde r\sim 0$ implying that it is spacelike. The future
trapping horizon $\tilde t^{(-)}_{\rm TH}(\tilde r)$
and other portion of past trapping horizon
$\tilde t^{(+)}_{\rm TH}(\tilde r)$  become timelike.  
In the limit $\tilde r\to 0$, the trapping horizons 
$\tilde t^{(\pm)}_{\rm TH}(\tilde r)$
have constant circumference radii $\tilde R_\pm$ as outside.

%It is important to note that 
%$\tilde R=[(\tilde r \tilde t+1)(\tilde r+1)^3]^{1/4}$ is divergent 
%for $-1<\tilde r<0$ in the limit $\tilde t\to -\infty $.
%This implies that infinity is present in the $\tilde r<0$
%region.  
%

%\bigskip\bigskip

\subsubsection{Constant $R$ surfaces}

The trapping horizons occur at 
$(\tilde \nabla_\mu  \tilde R)(\tilde \nabla^\mu \tilde R)=0$, i.e., 
the surfaces of $\tilde R={\rm constant}$ become null. 
Then the contours of circumference radius will help us to recognize the 
positions of trapping horizons in terms of circumference radius.

Solving Eq.~(\ref{R}) with respect to $\tilde t$, we obtain
\begin{eqnarray}
\tilde t={1\over \tilde r}\left[{\tilde R^4\over (1+\tilde r)^3}-1\right]
\,.
\end{eqnarray}
Taking the derivative of this equation
with fixing $\tilde R$, we find the relation between 
$\D \tilde t$ and $\D \tilde r$.  Inserting this to the metric~(\ref{4D_metric0}), 
the line element of  $\tilde R={\rm constant}$ surface  is given by
\begin{widetext}
\begin{eqnarray}
\D \tilde s^2=-{\tau^2(1+4\tilde r)^2\over \tilde r^2 
\tilde R^2 (1+\tilde r)^8}
\left(\tilde R^2-\tilde R_1^2\right)
\left(\tilde R^2+\tilde R_1^2\right)
\left(\tilde R^2-\tilde R_2^2\right)
\left(\tilde R^2+\tilde R_2^2\right)
\D \tilde r^2\,,
\end{eqnarray}
where
%---------------- R1 and R2 -----------------%
\begin{align}
\tilde R_1^2&=\tilde R^2 (\tilde t_{\rm TH}^{(+)}(\tilde r), \tilde r)= 
{(1+\tilde r)^4\over 2\tau |1+4\tilde r|}\left[1
+\sqrt{1+{4\tau^2(1+4\tilde r) \over (1+\tilde r)^4}}\right]\,,
\\
\tilde R_2^2&=\tilde R^2 (\tilde t_{\rm TH}^{(-)}(\tilde r), \tilde r)
={(1+\tilde r)^4\over 2\tau(1+4\tilde r)}\left[-1
+\sqrt{1+{4\tau^2(1+4\tilde r) \over (1+\tilde r)^4}}\right]
\,.
\end{align}
\end{widetext}
One finds that $\tilde R_1\to \tilde R_+$
and $\tilde R_2\to \tilde R_-$
in the limit $\tilde r \to 0$, and 
$\tilde R_1^2\to \infty$ and $\tilde R_2^2 \to  \tau $ as
 $\tilde r\to \infty $. 
It is notable that $\tilde R_1$ and $\tilde R_2$ are not independent but
fulfills the constraint 
\begin{align}
 \tau =\frac{\tilde R_1^2 \tilde R_2^2}
{\tilde R_1^2-\tilde R_2^2}\,.
\end{align}

From  $\tilde R_1^2>\tilde R_2^2>0>-\tilde R_2^2>-\tilde R_1^2$,
we conclude that
if $\tilde R^2>\tilde R_1^2$ or $\tilde R^2<\tilde R_2^2$, 
the $\tilde R={\rm constant}$ curves are  
timelike, while they are spacelike when $\tilde R_2^2<\tilde R^2<\tilde R_1^2$.
In the region where $\tilde R={\rm constant}$ curves are  
timelike (spacelike), a metric sphere is untrapped (trapped). 

\begin{figure}[t]
\includegraphics[width=7cm]{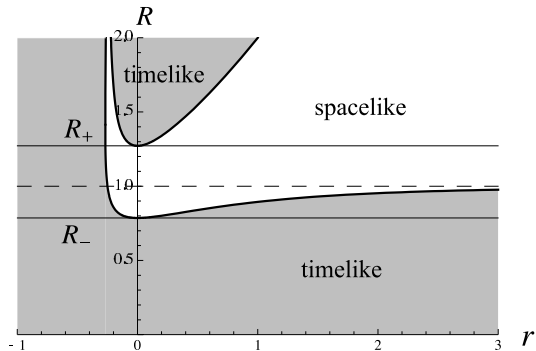}
\caption{Signature of $\tilde R={\rm constant}$ surfaces
 for $\tau =1$. 
$\tilde R={\rm constant}$ curve is timelike (grey region) and 
spacelike (white region), which are separated by 
trapping horizons $\tilde t^{(\pm)}_{\rm TH}$}
\label{fig:constantR}
\end{figure}

We sketch in Figure~\ref{fig:constantR} 
the region where $\tilde R={\rm constant}$ curves are timelike (grey
region), and spacelike (white region).  
As $\tilde r\to 0$, we find that $\tilde R_\pm$ are null
surfaces.

%%%%%%%%%%%%%%%%%%%%%%%%%%%%%%%%%%%%%%%%%%%%
\subsection{Event horizons}
%%%%%%%%%%%%%%%%%%%%%%%%%%%%%%%%%%%%%%%%%%%%
\label{sec:EH}

Let us proceed to demonstrate the structure of the future and past event horizons, utilizing
several results obtained in the previous sections.

Before embarking on this programme, let us recapitulate basic features
of event horizons. The future event horizon is defined by a future
boundary of the causal past of future null infinity. 
The past event horizon is similarly defined by interchanging the r\^ole
of future and past. 
These event horizons are by definition the achronal 
3D null surfaces. In addition, the future (past) event horizon 
is generated by null geodesic generators which have no future (past) endpoint in 
$(M, g_{\mu \nu })$~\cite{HE}.

If the spacetime is stationary, the black hole event horizon must be 
a Killing horizon~\cite{HE} (see~\cite{Hollands:2008wn} for the 
degenerate case). This theorem is considerably significant because
it enables us to identify the locus of event horizon simply from the 
spacetime symmetry. It is {\it a priori} no interrelationship
between the event horizon and the  Killing horizon. 
Such a lucky consequence is exceptional to
the stationary case. 

In the dynamical case, on the other hand, we have no 
specific guidelines for identifying the future event horizon
but to evolve the spacetime into the infinite future.\footnote{
The Sultana-Dyer solution~\cite{SultanaDyer} is an 
exceptional instance: it is conformal to the
Schwarzschild  metric  hence its causal picture is extracted 
in a simple fashion~\cite{Saida:2007ru}.  
}  
Nevertheless, 
we can say, regardless of this adversity, that the 
black hole event horizon has to cover the trapped surfaces, 
provided the outside region of a black hole is sufficiently
well-behaved~\cite{HE}.  
Inspecting that the spacetime~(\ref{4D_metric0}) appears to have 
a good behavior at least for $\tilde r>0$ and that the 
$\tilde \theta_\pm =0$ surfaces comprise null 
surfaces~(\ref{R_plusminus}) in the limit $\tilde r\to 0$
and $\tilde t\to \pm \infty $, it may be reasonable to
consider these null surfaces as the possible 
candidates of black and white hole horizons. Analyzing 
the near-horizon geometry and behaviors of null geodesics, 
we shall see below that this expectation is indeed true.

\subsubsection{Near horizon geometry}

From the behavior of trapped surfaces, 
we can deduce that the null surface $\tilde r=0$
is a plausible horizon candidate. 
We shall scrutinize the structure of this surface 
in detail.  

We first look at the ``throat'' geometry~(\ref{near_horizon}). 
Taking the fiducial time $\tilde t=1$ for simplicity (the same
conclusion is derived for any value of finite $\tilde t$), the 
proper distance $\tilde s$ from  the spacetime point 
$(1, \tilde r, \tilde\theta , \tilde\phi)$ to 
$(1, \tilde r=0, \tilde \theta , \tilde\phi)$ 
is given by 
%--------------- proper distance to throat ---------------%
\begin{align}
\tilde s&= \lim_{\tilde r_0 \to 0 } \int ^{\tilde r}_{\tilde r_0} 
\sqrt{\tilde g_{\tilde r\tilde r} (1, \tilde r)} \D \tilde r  
\nonumber \\
&=\lim_{\tilde r_0 \to 0 } \biggl[\tilde r +\ln \tilde
 r\biggl]_{\tilde r_0}^{\tilde r} \to
 \infty \,.
\label{distance}
\end{align}
This implies that the point $\tilde r=0$  corresponds 
not to the regular origin of polar coordinates but to  
``spatial infinity,'' as in the extremal RN spacetime. 
In the extremal RN case, the future (past) event horizon is a null surface 
generated by $\tilde r=0$ and $\tilde t= \infty $ ($t=-\infty $), with its
infinite ``throat'' at
$\tilde r=0$ with $\tilde t$ being finite.  
Analogously, 
the event horizon of present spacetime,   
if it exists, should have $\tilde r=0$ with finite 
$\tilde t$ as its ``throat.''
So we deduce that the only candidate of event horizons
in the present spacetime is $\tilde r=0$ and $\tilde t=\pm \infty$.

Therefore, the most convenient way to see the 
structure of these candidate horizons is to take 
the {\it  near-horizon limit},  defined by 
%---------------- near-horizon limit  ------------------%
\begin{align}
\tilde  t\ \to \frac{\tilde t}{\epsilon }, \qquad 
 \tilde r\to \epsilon \tilde r, \qquad \epsilon \to 0\,,
\label{rescaling}
\end{align}
where $\epsilon $ is a positive constant. 
Other conceivable limits fail to produce any sensible results.   
After the rescaling, the coordinate ranges of $\tilde t$ and $\tilde r$ 
are free from a restriction other than $\tilde r>0$.  
Taking the limit (\ref{rescaling}) in Eq.~(\ref{4D_metric0}), 
one obtains 
the near-horizon metric: 
%----------------- NH geometry ---------------------% 
\begin{align}
\D \tilde s_{\rm NH}^2 =&
-{\tau^2 \tilde r^2 (1+\tilde t \tilde r)^{-1/2}} 
 \D \tilde t^2
\nonumber \\
&
+\tilde r^{-2}{(1+\tilde t\tilde r)^{1/2}} \left(\D
 \tilde r ^2+\tilde r^2 \D \Omega_2^2 \right)\,,
\label{NHmetric}
\end{align}
which does not involve the parameter $\epsilon $. 
As a direct consequence of (\ref{rescaling}), the near-horizon metric~(\ref{NHmetric})
is invariant under the flow
%----------------- Killing vector  -----------------%
\begin{align}
 \xi^\mu =\tilde t\left(\frac{\partial }{\partial \tilde t}\right)^\mu 
-\tilde r\left(\frac{\partial }{\partial \tilde r}\right)^\mu \,.
\label{horizonKilling}
\end{align}
Namely, $\xi^\mu $ is a Killing vector in the
spacetime~(\ref{NHmetric}).

Changing the coordinate $\tilde r $ to the 
circumference radius 
$\tilde R =  \left(1+ \tilde t\tilde r \right)^{1/4}$, 
the near-horizon metric~(\ref{NHmetric}) transforms into  
%------------- NH metric tR ----------------%
\begin{align}
 \D \tilde s_{\rm NH}^2
=& -f(\tilde R){\D \tilde t^2\over \tilde t^2 \tilde R^2}
-8{\tilde R^5\over \tilde t(\tilde R^4-1)}\D \tilde t \D \tilde R 
\nonumber \\
&+{16\tilde R^8\over (\tilde R^4-1)^2}\D \tilde R^2+\tilde R^2 \D \Omega_2^2\,,
\label{NH_metric_tR}
\end{align}
where
%------------------ f(R)  -----------------% 
\begin{align}
 f(\tilde R):=&\tau^2(\tilde R^4-1)^2-\tilde R^4 \nonumber \\
=& \tau^2 
(\tilde R^4-\tilde R_+^4)(\tilde R^4-\tilde R_-^4 )
\,.
\label{4D_metric4}
\end{align}
Here, $\tilde R_+$ and $\tilde R_-$ have been defined in Eq.~(\ref{R_plusminus}).
In the coordinates~(\ref{NH_metric_tR}), we have
$\xi^\mu =\tilde t(\partial /\partial \tilde t)^\mu $. 
Apart from  $\tilde R=0$ (which is
indeed a curvature singularity) and the points at $\theta=0, \pi $  (which
are north and south poles of 2-sphere),  
there appear additional coordinate singularities 
at $\tilde t=0$ and $\tilde R=1$ in the metric~(\ref{NH_metric_tR}).

Although the metric~(\ref{NH_metric_tR}) is time-dependent, 
we can eliminate the time-dependence of the metric~(\ref{NH_metric_tR})
by changing to the time slice,
%------------------ eta  -----------------% 
\begin{eqnarray}
\eta_\pm := \ln (\pm \tilde t)\,,~~~
{\rm for }~~ \tilde t\gtrless 0,
\end{eqnarray} 
in terms of which the Killing vector is written as
 $\xi^\mu =(\partial /\partial \eta _\pm)^\mu $ and 
the near-horizon metric (\ref{NH_metric_tR}) is given
by 
\begin{align}
\D \tilde s^2_{\rm NH}=&-{f(\tilde R)\over \tilde R^2}\left[\D \eta_\pm 
+{4\tilde R^7\over (\tilde R^4-1)f(\tilde R)}\D \tilde R\right]^2
\nonumber \\&
+{16\tau^2 \tilde R^8\over f(\tilde R)}\D \tilde R^2+\tilde R^2 \D
 \Omega_2^2 
\,.
\label{4D_metric3}
\end{align} 
The sign of $\eta_\pm$ has been chosen in such a way that
 $\eta_+ ~(\eta_-) $ increases (decreases) as $\tilde t$ increases.

Performing a further coordinate transformation, 
\begin{align}
 \tilde T _\pm = \eta_\pm +
\int ^{\tilde R}{4\tilde R^7\over (\tilde R^4-1)f(\tilde R)}\D \tilde
 R\,, 
\label{tilde_T}
\end{align}
the near-horizon metric~(\ref{4D_metric3}) is brought into a 
familiar form,
%------------- static NH black hole -----------%
\begin{align}
 \D  \tilde s^2_{\rm NH} =
-{f(\tilde R)\over \tilde R^2} \D \tilde T^2_\pm 
+{16\tau^2 \tilde R^8\over f(\tilde R)}\D \tilde R^2+\tilde R^2 \D
 \Omega_2^2 \,.
\label{NHmetric2}
\end{align}
This metric describes a static black hole whose  
horizons occur where the lapse function vanishes $f(\tilde R)=0$,
i.e., where the Killing field 
$\xi^\mu =(\partial /\partial \tilde T_\pm)^\mu $ becomes null.
The condition $f(\tilde R)=0$ gives two roots 
$\tilde R=\tilde R_\pm$ given by  Eq.~(\ref{R_plusminus}), 
which coincide with the trapping horizons in the $\tilde r\to 0$
limit taken. Thus, we conclude that the null surfaces 
$\tilde R=\tilde R_\pm$ in the original spacetime are locally isometric
to the Killing horizons in the static spacetime (\ref{NHmetric2}).

Reminding the fact that the outside domain of the original 
spacetime~(\ref{4D_metric0}) is highly dynamical and 
hence is lack of 
non-spacelike Killing field,  
it comes out a novel surprise for us  that 
the near-horizon metric~(\ref{NHmetric2}) permits the unexpected 
symmetry~(\ref{horizonKilling}). 
Observe that 
the vector field (\ref{horizonKilling}) satisfies the Killing equation
in the original spacetime (\ref{4D_metric0}) 
{\it only at the horizon $\tilde R=\tilde R_\pm$}. 
This may be ascribed to the fact that the
11D solution is supersymmetric if all branes are at rest. 
The supersymmetry does not allow energy inflow, 
consistent with the property that the Killing horizon
is totally geodesic.  This may be clear by  
considering the Raychaudhuri equation:  
$T_{\mu \nu }\xi^\mu \xi^\nu \to 0$ is indeed satisfied in the 
limit (\ref{r0limit1}) or (\ref{r0limit2}).
As far as the authors know, 
this is a first realization of 
asymptotic symmetry appearance at the black-hole horizon
under the dynamical circumstance.

%------------------ Static BH analysis ----------------%

\begin{figure}[t]
\begin{center}
\includegraphics[width=5cm]{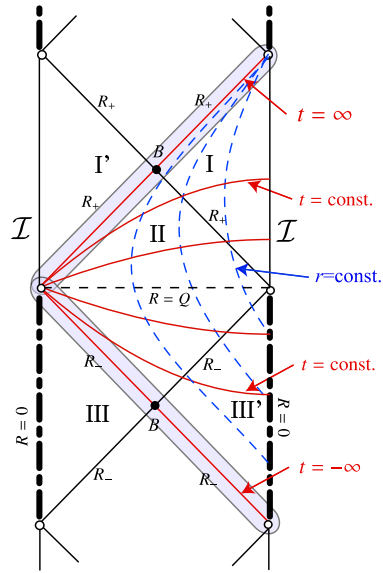}
\caption{Conformal diagram of a maximally extended 
near-horizon metric~(\ref{NHmetric2}). The black hole
has outer and inner horizons, whose radii are respectively 
given by $\tilde R_+$ and $\tilde R_-$. Infinity consists of
a timelike surface denoted by $\mathcal I$. 
The white circles mark points at infinity ($i^0$, $i^\pm$ or the ``throat'') and should not be regarded as
spacetime events. The filled circles $B$ represent bifurcation surfaces 
{\it for the metric}~(\ref{NHmetric2}) at which 
$\xi^\mu =(\partial/\partial T_\pm)^\mu $ vanishes, i.e., $B$'s are the fixed points
 under this isometry.  
Thick red lines correspond to $\tilde t ={\rm constant}$ surfaces, and
blue dotted lines denote $\tilde r={\rm constant}$ surfaces. $\tilde R=1$ is a coordinate
 singularity for the metric~(\ref{NH_metric_tR}) corresponding to
 $\tilde t=0$. 
The shaded regions 
approximate our dynamical metric~(\ref{4D_metric0}) in the neighborhood
 of horizons. }
\label{PD_static}
\end{center}
\end{figure}

Let us devote some space here  to discuss the near-horizon static
metric~(\ref{NHmetric2}) in more detail.  In this limit, the 
dilaton~(\ref{Phi}) and the Maxwell fields~(\ref{form_field}) are 
reduced to  
\begin{align}
 \kappa \Phi =\sqrt{6} \ln \tilde R\,, 
\label{NHdilaton}
\end{align}
and
\begin{align}
 \kappa \tilde F^{(T)} &= -4\sqrt{2\pi}\tau\tilde R^{-5}~
\D \tilde T\wedge \D \tilde R\,, \nonumber \\
 \kappa \tilde F^{(S)} &= -{4\sqrt{2\pi}\tau}{\tilde R^3}~
\D \tilde T\wedge \D \tilde R\,.
\label{NHMaxwell}
\end{align}
We can confirm Eqs.~(\ref{NHmetric2}), (\ref{NHdilaton}) and (\ref{NHMaxwell})
still satisfy the original field equations~(\ref{4D_Einstein}), (\ref{eq_dilaton}) and
(\ref{eq_U(1)}), which justifies  that the near-horizon limit (\ref{rescaling})
is well-defined. 
It is obvious that 
the near-horizon metric (\ref{NHmetric2}) describes a static black hole, 
whose asymptotic structure is neither flat nor AdS.  
Such an unusual asymptotic structure is,
 however, that one commonly encounters in 
Einstein-Maxwell-dilaton gravity (see e.g.,~\cite{Chan:1995fr,Yazadjiev:2005du}). 
Albeit this peculiar asymptotics, it is easy to find that 
the causal structure is akin to that of the nonextremal RN-AdS solution
(see Figure~\ref{PD_static}).
The spatial infinity $\tilde R\to \infty $ consists of a timelike
boundary $\mathcal I$, 
a timelike singularity resides at the center $\tilde R=0$, 
and the two distinct outer and inner horizons $\tilde R_\pm$ 
of black hole and white hole arise.

Since $\tilde R_+$ is strictly larger than $\tilde R_-$
for finite $\tau $
[see Eq.~(\ref{Qtau})], we find that the horizons are
not degenerate. 
From the general formula 
$\kappa^2_{\pm } = \mp (1/2)(\nabla_\mu \xi_\nu )\nabla^\mu \xi^\nu $,
one obtains the surface gravities of these
horizons,\footnotetext{It deserves to mention that the values of surface
gravities are sensitive to the norm of the generator
$\xi^\mu $ of the Killing horizon. 
The spacetime~(\ref{NHmetric2}) is not asymptotically flat, 
hence there exists no meaningful way to fix the values. (In the
asymptotically flat case, we usually require at infinity $\xi_\mu\xi^\mu \to -1$,
which eliminates the ambiguity). However, this ambiguity   
causes no harm in our present discussion. The important point to note
here is that they take nonvanishing positive values. We will revisit
this issue when we discuss thermodynamics in the ensuing section.}
\begin{align}
\kappa _\pm =\pm \frac{f' (\tilde R_\pm ) }{8\tau \tilde R_\pm ^5}
=\frac{\sqrt{1+4\tau^2 }}{\sqrt{1+4\tau^2}\pm 1}\,,
\label{kappa_pm}
\end{align}
where $\kappa_+$ and $\kappa_- $ are the surface gravity of
the outer and inner event horizons, respectively. 
The surface gravity of the Killing horizon is constant 
over the horizon, illustrating the equilibrium state.  
The nonvanishing surface gravity might be seemingly puzzling, 
in light of the situation that the 11D solution was ``maximally
charged'' because of the supersymmetry in the static limit.

Introduce the null coordinates $u_\pm$ by  
\begin{align}
  u_\pm &=\tilde T_\pm -\int^{\tilde R}\frac{4\tau \tilde R^5}{f(\tilde
 R)}\D \tilde R \nonumber \\&= 
%\ln \left[\pm \frac{\tilde t|\tilde
% R^2+ \tilde R_+^2|^{1/\kappa _+}|\tilde R^2- \tilde
% R_-^2|^{1/\kappa_- }}{|\tilde R^4-1|}\right]
\ln \left[\pm \tilde t|\tilde R^4-1|^{-1}|\tilde
 R^2+ \tilde R_+^2|^{1/\kappa _+}|\tilde R^2- \tilde
 R_-^2|^{1/\kappa_- }\right], 
\end{align}
where the integration follows from a direct calculation by  
Eqs.~(\ref{4D_metric4}), (\ref{tilde_T}) and (\ref{kappa_pm}). 
The coordinates $u_\pm$ are well-defined at $\tilde R=\tilde R_+$. 
Using $u_\pm$, 
the metric~(\ref{NHmetric2}) is translated into the single null form, 
%----------------- single null ------------------%
\begin{align}
 \D \tilde s^2_{\rm NH} =-\frac{f(\tilde R)}{\tilde R^2 } \D u_\pm ^2- 2 \D
 u_\pm \D \tilde R +\tilde R^2 \D \Omega_2^2\,.
\label{NHmetric_null} 
\end{align}
Let us consider the plus-coordinate in Eq.~(\ref{NHmetric_null}) and 
discuss the outer white hole horizon (the boundary of I' and II). 
It is immediate to find that the null generator of the Killing horizon  
$\xi^\mu $
is expressed in this coordinates as 
$\xi^\mu =(\partial/\partial u_+)^\mu $.
$u _+$ is the Killing parameter ($\xi^\mu\nabla_\mu u_+  =1$) of null geodesic generators. 
It then follows that the renormalized tangent vector
%-------------- affine geodesic ----------------%  
\begin{align}
 k^\mu =\frac{1}{\kappa_+ }e^{\kappa_+  u_+  }\left(
\frac{\partial }{\partial u_+  }\right)^\mu ,
\end{align}
is affinely parametrized, i.e., 
$k^\mu $ satisfies the zero-acceleration geodesic equation 
$k^\nu \nabla_\nu k^\mu =0$ on the horizon. (Note that
we are considering a white hole so that 
$\xi ^\nu \nabla_\nu \xi^\mu =-\kappa_+ \xi^\mu $.)  
This means that the affine parameter $\lambda $ 
[$k^\mu =(\partial /\partial \lambda )^\mu $] is related to the
Killing parameter $u_+$ as
%---------------  u_+ and lambda --------------%
\begin{align}
\lambda  =-e^{-\kappa_+ u_+}.
\end{align} 
This manifests that the affine parameter $\lambda $ 
values from $-\infty $  to $0 $ as $u_+$ ranges from 
$-\infty $ to $\infty $, implying the bifurcation surface--a closed
surface at which $\xi^\mu $  vanishes.
This can be verified by noticing 
$\xi^\mu =(\partial/\partial \tilde T_+)=\kappa_+ \lambda (\partial/\partial \lambda )^\mu \to 0$ 
as $\lambda \to 0$. 
The similar argument goes through to $u_-$.  
Therefore, the nondegenerate Killing horizon
(i.e., a Killing horizon with nonvanishing surface gravity)  
is incomplete either into the past or future. 
This is a general consequence of a Killing horizon~\cite{Carter}.

However, the above discussion does not mean that the horizon in our 
original spacetime  is the bifurcate Killing horizon.   
This is consistent with results in~\cite{Racz} which
asserts that the nondegenerate Killing horizon is the bifurcate Killing
horizon. In their proof, it is assumed that the horizon is smooth 
($C^\infty$-class), while the horizon 
in the present case is only finite times differentiable 
($C^k$-class with $k$ finite) (see Section~\ref{sec:asysol} below).

To summarize this subsection, 
we can expect that the original spacetime~(\ref{4D_metric0}) would have  
the  future-event horizon at $\tilde r\to 0$ with $\tilde t\to \infty $, and 
the past-event horizon at $\tilde r\to 0$ with $\tilde t\to -\infty $.
We have  found that the point at $\tilde r=0$ with $\tilde t$
being finite corresponds to ``throat
infinity'' just as that of the extreme RN spacetime, at which future and  past
event horizons should intersect. 
In the neighborhood of these horizon candidates, 
the spacetime~(\ref{4D_metric0}) is approximated by
the near-horizon geometry~(\ref{NHmetric2}) with Killing horizons, in 
which several portions of Killing horizons with radii $\tilde R_\pm$ appear. 
What portion of Killing horizons in Figure~\ref{PD_static} does it correspond to 
the horizon in our original spacetime? The answer is obvious:  
the ``white hole portion'' [grey-colored line segment encompassing 
blocks I, II and III' in Figure~\ref{PD_static}] only satisfies  
the above criteria.

%%%%%%%%%%%%%%%%%%%%%%%%%%%%%%%%%%%%%%%%%%%%
\subsubsection{Null geodesics}
%%%%%%%%%%%%%%%%%%%%%%%%%%%%%%%%%%%%%%%%%%%%

We marked out $\tilde R_+$ as a black hole horizon
in the spacetime~(\ref{4D_metric0}). To conclude this 
more rigorously,  
we face up the problem of solving  geodesic motions. 
Since the present spacetime is spherically symmetric, 
it suffices us to focus on radial null geodesics
to argue causal structures.
Although examinations of nonradial and/or timelike geodesic motions are 
important issues in order to clarify  the detailed physical properties of the 
solution,  we will not discuss these since behaviors of radial null
geodesics are sufficient to determine the causal structure.

The radial null geodesic equations are governed by 
%------------- Radial null geodesics  ------------%
\begin{align}
 \ddot{\tilde t}-\frac{1}{4H_T}\dot{\tilde t}^2+\frac{(H_S+3H_T)}{2\tilde
 r^2H_TH_S}\dot{\tilde t}\dot{\tilde r}+\frac{H_S^3}{4\tau^2
 }\dot{\tilde r}^2 &=0\,, 
\label{geo1}\\
\ddot{\tilde r}+\frac{\tau^2 (H_S+3H_T)}{4\tilde
 r^2H_T^2H_S^4}\dot{\tilde  t}^2+\frac 1{2 H_T}\dot{\tilde 
 t}\dot{\tilde r}&
\nonumber \\
-\frac{(H_S+3H_T)}{4\tilde r^2H_TH_S}\dot{\tilde r}^2&=0\,,\label{geo2}\\
-\tau^2 \dot{\tilde t}^2+H_TH_S^3 \dot{\tilde r}^2&=0\,,\label{geo3}
\end{align}
where the dot denotes a differentiation with respect to an affine
parameter $\lambda $. 
These equations are combined to give
%------------- Null geodesics  ------------%
\begin{align}
 \ddot{\tilde t} \pm \frac{ \tau (H_S+3H_T)}{2\tilde r^2(H_T^3H_S^5)^{1/2}}
\dot{\tilde t}^2&=0\,,\label{geo4} \\
\ddot{\tilde r}\pm \frac{(H_TH_S^3)^{1/2}}{2\tau H_T}\dot{\tilde
 r}^2&=0\,,
\label{geo5}
\end{align}
where the plus (minis) sign refers to the outgoing (ingoing) geodesics. 
Unfortunately, the radial null geodesics do not appear to 
admit a first integral other than Eq. (\ref{geo3}), 
so it is not amendable to analytic study. Instead, 
we try to solve numerically Eqs.~(\ref{geo4}) and (\ref{geo5}) subjected to
the initial constraint (\ref{geo3}).  
Making use of the degrees of
freedom of the affine parameter $\lambda \to a \lambda+b$, 
we are able to choose $\lambda=0$ at the starting point of the geodesics and set 
$\dot{\tilde t}(0) $ at any values we wish. 
Fixing the orientation of future-directed geodesics to be $\dot{\tilde t}>0$, and 
past-directed to be $\dot{\tilde t}<0$, 
we choose $\dot{\tilde t}(0) \equiv 1~(-1)$
for future (past) directed radial null geodesics without loss of
 generality.
Hence the residual freedoms that distinguish different geodesics 
are two, corresponding to the initial values 
[$\tilde t(0)$ and $\tilde r(0)$] for each $\tau>0 $.

Let us begin our consideration by 
the geodesics in the outside region $\tilde r>0$. 
Taking the representative spacetime events $p_I~(I=1,2,3)$ such that 
$ \tilde t_1>\tilde t_{\rm TH}^{(+)}$, 
$\tilde t_{\rm TH}^{(-)}<\tilde t_2< \tilde t_{\rm TH}^{(+)}$ and
$\tilde t_3<\tilde t_{\rm TH}^{(-)}$ where $\tilde t_I\equiv \tilde
t|_{p_I}$
(see Figure~\ref{fig:TH}), 
we have examined behaviors of geodesics starting from
$\tilde t=\tilde t_I$.  
We call the geodesics emanating from the event $p_I$ as
Class-$I$. Since $\tilde R_1>\tilde R_+$ ($\tilde R_2<\tilde R_-$),
Class-1 (Class-3) geodesics initially have a circumference radius
larger (smaller) than $\tilde R_+ $ ($\tilde R_-$). 

We depict several typical geodesic curves 
emanating from $(\tilde t_I, \tilde r(0)=1)$ 
for $\tau=1$ in Figure~\ref{geodesics_figures}. 
This is a representative figure for $\tau<\tau_{\rm crit}$. 
Qualitative behavior of geodesics seems not 
so sensitive to the initial radial position $\tilde r(0)$.
The numerical results are summarized as follows:

\begin{widetext}
\begin{center}
\begin{figure}[h]
\includegraphics[width=9cm]{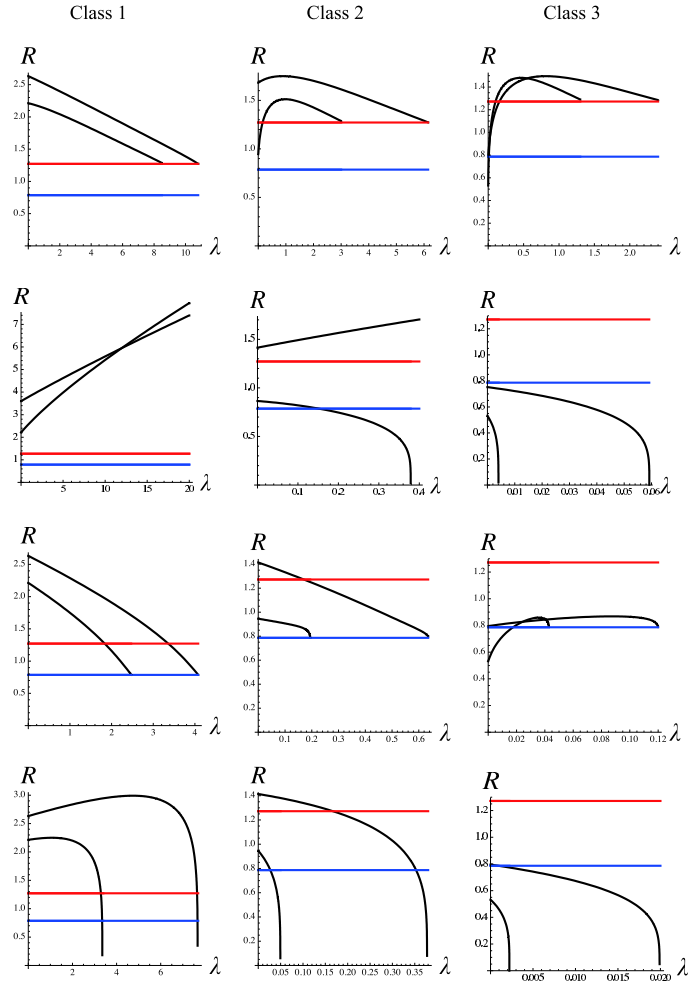}
\caption{Radial null geodesics in the outside region ($\tilde r>0$)
emanating from $\tilde t(0)=\tilde t_I$ and $\tilde r(0)=1$ for
$\tau =1$. The diagrams in the top, upper  middle, lower middle, bottom
rows correspond to future-directed ingoing null geodesics, 
future-directed outgoing null geodesics, past-directed ingoing null
 geodesics and past-directed outgoing null geodesics.
The red and blue lines denote 
$\tilde R_+$ and $\tilde R_-$, respectively. }
\label{geodesics_figures}
\end{figure}
\end{center}
\end{widetext}

\bigskip\noindent
$\bullet$
{\it Future-directed ingoing null geodesics:}
Class 1 geodesics monotonically decrease the circumference radius 
and arrive at $\tilde R_+$ within a finite affine time. 
Class 2 geodesics first increase circumference radius 
since they are originally in the trapped region $\tilde \theta_->0$. But 
they always move across $\tilde t^{(+)}_{\rm TH}$, and finally reach 
$\tilde R_+$ with decreasing area.    
The qualitative behavior of Class 3 are the same as that of Class 2, 
except that Class 3 geodesics always cross $\tilde R=\tilde R_+$ twice, and if  
$\tilde t_3$ is sufficiently close to $\tilde t_{s}$, they may cross 
$\tilde R_-$.
All classes of geodesics have infinite redshift
$\tilde t(\lambda )\to +\infty $ when they finally arrive at $\tilde R_+$.

\bigskip\noindent
$\bullet$
{\it Future-directed outgoing null geodesics:}
Class 1 geodesics necessarily go out to
infinity $\tilde R\to \infty $. 
Class 2 geodesics may extend out to infinity or 
arrive at the singularity $\tilde t=\tilde t_s$ if 
$\tilde t_2$ is small.  
Class 3 geodesics  inevitably plunge into the 
singularity $\tilde t=\tilde t_{s}$ within a finite affine time.

\bigskip\noindent
$\bullet$
{\it Past-directed ingoing null geodesics:}
Class 1 geodesics  originate from $\tilde R>\tilde R_+$ and their radii
monotonically decrease toward $\tilde R_-$. Class 2 geodesics
have qualitatively  the same behavior. 
Class 3 geodesics start from $\tilde R<\tilde R_-$ with increasing area,  
then cross $\tilde R_-$ (with finite $\dot{\tilde t}$), attain the maximum radius and 
get back to $\tilde R_-$ again with undergoing infinite blueshift.

\bigskip\noindent
$\bullet$
{\it Past-directed outgoing null geodesics:}
Class 1 geodesics may initially increase the area, but  
all geodesics unavoidably terminate into the singularity 
$\tilde t=\tilde t_s$ 
within a finite affine time.

%\bigskip
From these results, we conclude that the null surfaces $\tilde R=\tilde R_\pm$
locate within a finite affine time from outside spacetime events. 
Behaviors of future-directed outgoing null rays of Class 2 geodesics 
imply that there exists a critical null curve $\tilde t=\tilde t_*(\tilde r)$
such that outgoing rays emanating from $\tilde t>\tilde t_*$ can get to
infinity, whereas outgoing rays emanating from $\tilde t<\tilde t_*$
fall into the singularity.

 \begin{widetext}
\begin{center}
\begin{figure}[h]
\includegraphics[width=9cm]{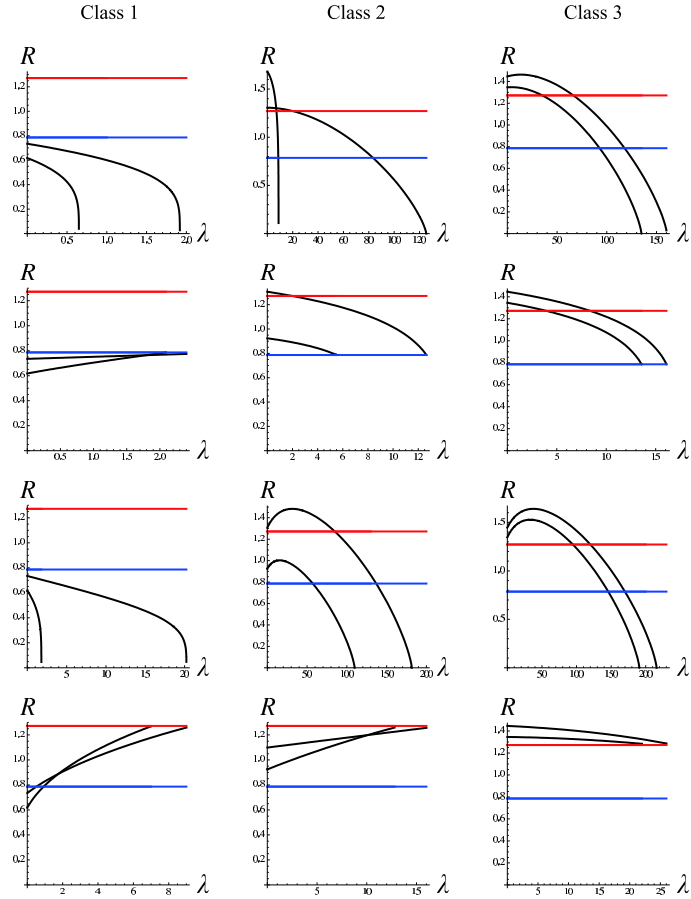}
\caption{Radial null geodesics in the inside region ($\tilde r<0$)
emanating from $\tilde t(0)=\tilde t_I$ and $\tilde r(0)=-1/10$ for
$\tau =1$. 
The diagrams in the top, upper  middle, lower middle, bottom
rows correspond to future-directed ingoing null geodesics, 
future-directed outgoing null geodesics, past-directed ingoing null
 geodesics and past-directed outgoing null geodesics.
The red and blue lines denote 
$\tilde R_+$ and $\tilde R_-$, respectively.
}
\label{geodesics_figures2}
\end{figure}
\end{center}
\end{widetext}

%--------------- Geodesics  r<0 ----------------%
Let us discuss next the geodesics inside the horizon. 
We call Class 1 as 
$\tilde t >\tilde t_1>\tilde t^{(-)}_{\rm TH}$, 
Class 2 as 
$\tilde t^{(-)}_{\rm TH} >\tilde t_2>\tilde t^{(+)}_{\rm TH}$ and
Class 3 as $\tilde t^{(+)}_{\rm TH}>\tilde t_3$ (see
Figure~\ref{TH_negativer}). 
Figure~\ref{geodesics_figures2} plots the geodesic curves emanating from
the spacetime event $(t_I, \tilde r(0)=-1/10)$ with 
$\dot{\tilde t}(0)=\pm 1$. 
Geodesics starting from $\tilde r(0)<\tilde r_0$ show the same 
behavior as Class 1.  
The result is:

\bigskip
\noindent
$\bullet$
{\it Future-directed ingoing null geodesics:}
Class 1 geodesics and Class 2 geodesics starting from 
not so small $\tilde t_2$ eventually fall into the 
singularity $\tilde t=\tilde t_s$. Class 2 geodesics with
$\tilde t_2$ close to $\tilde t^{(+)}_{\rm TH}(\tilde r)$ 
 fall into the
singularity $\tilde r=-1$ within a finite affine time. 
Class 3 geodesics initially increase area, but they
eventually plunge into the singularity at $\tilde r=-1$.

\bigskip\noindent
$\bullet$
{\it Future-directed outgoing null geodesics:}
Class 1 geodesics increase area and approach
$\tilde R=\tilde R_-$ with infinite redshift 
$\tilde t(\lambda ) \to + \infty $. 
Class 2 and Class 3 geodesics decrease area and 
arrive at $\tilde R=\tilde R_-$ with infinite redshift 
$\tilde t(\lambda ) \to + \infty $.

\bigskip\noindent
$\bullet$
{\it Past-directed ingoing null geodesics:}
Class 1 geodesics monotonically reduce area 
and fall into the singularity $\tilde t=\tilde t_s $. 
Class 2 and Class 3 geodesics initially grow the area and then turn to
decrease, and finally fall into the singularity $\tilde r=-1$.

\bigskip\noindent
$\bullet$
{\it Past-directed outgoing null geodesics:}
All geodesics eventually approach $\tilde R=\tilde R_+$ 
with infinite blueshift $\tilde t (\lambda )\to -\infty $.

\bigskip
From these results, the null surface $\tilde R=\tilde R_-$ 
has an ingoing null structure, analogous to the white hole horizon or
the black hole inner horizon.

\subsubsection{Asymptotic solutions of geodesics}
\label{sec:asysol}

We have numerically established that the 
radial null geodesics are incomplete
at the null surfaces $\tilde R=\tilde R_\pm$. 
We shall look into the asymptotic geodesic solutions
and discuss further the horizon structure.

If a null geodesic is known as
 $\tilde t=\tilde t(\tilde r)$, the affine parameter
$\lambda $ is obtained by a simple quadrature~\cite{BHKT}
\begin{align}
 \lambda =\int \D \tilde r \exp\left[
\pm \int U[\tilde t(\tilde r'), \tilde r']\D r' \right]\,,
\label{affine_sol}
\end{align}
where we have used the shorthand notation 
\begin{align}
 U = \frac{(H_TH_S^3)^{1/2}}{2\tau H_T}\,.
\end{align}
In the event horizon limit (\ref{r0limit1}), we have
\begin{align}
 U~ \to ~ \frac{1}{\tilde r (1+\sqrt{1+4\tau^2})}\,.
\end{align}
Substituting this into Eq.~(\ref{affine_sol}) and solving 
with respect to the radial coordinate, 
we obtain the asymptotic solution of the future-directed 
null geodesic around the horizon $\tilde R=\tilde R_+$ as
\begin{align}
\tilde r=c_1^{(+)} (\lambda -\lambda_+)^{1/\kappa_+}\,, \quad
\tilde t=c_2^{(+)} (\lambda -\lambda_+)^{-1/\kappa_+}\,,
\label{asysol_futureEH}
\end{align}
where $\lambda_+$ corresponds to the arrival time
for the geodesics at the horizon,  and $c_1^{(+)}$ and $c_2^{(+)}$ are 
positive constants
satisfying $c_1^{(+)}c_2^{(+)} =(1+\sqrt{1+4\tau^2})/(2\tau^2)$.
$\kappa _+$ has been given in Eq.~(\ref{kappa_pm}).
We can find from Eq.~(\ref{asysol_futureEH}) that the radial geodesics
indeed reach the horizon within a finite affine time~\cite{BHKT}.  
Equation (\ref{asysol_futureEH}) implies that 
$\tilde r$ and $\tilde t$ are not smooth functions of $\lambda $ 
[note that $1/\kappa_+$ never takes an integer].

Similarly, we obtain 
\begin{align}
 \tilde r=c_1^{(-)} (\lambda -\lambda_-)^{1/\kappa_-}\,, \quad
\tilde t=c_2^{(-)} (\lambda -\lambda_-)^{-1/\kappa_-}\,,
\end{align}
for an outgoing null geodesic near the horizon $\tilde R=\tilde R_-$.
Constants $c_1^{(-)}$ and $c_2^{(-)}$ satisfy 
 $c_1^{(-)}c_2^{(-)} =(1-\sqrt{1+4\tau^2})/(2\tau^2)$.

%%%%%%%%%%%%%%%%%%%%%%%%%%%%%%%%%%%%%%%%%%%%
\subsection{Carter-Penrose diagram}
%%%%%%%%%%%%%%%%%%%%%%%%%%%%%%%%%%%%%%%%%%%%

We are now in a position to discuss 
global causal structures of spacetime, by assembling 
considerations hitherto obtained.  
The optimal way to appreciate the large scale causal structure
is to draw the Carter-Penrose conformal diagram, 
which enables us to visually capture the global light-cone fabric. 
We first notice the followings:

\bigskip
\begin{itemize}
\item[(i)] The only candidate of future and past event horizons are $\tilde r=0$ and 
$\tilde t\to \pm \infty$, which are joined at the ``throat'' at $\tilde r=0$ and
$\tilde t$ being finite.

\item[(ii)] The near-horizon geometry of the event horizons is locally isometric to
       that of the static black hole~(\ref{NHmetric2}). The white-hole
	    portion corresponds to the horizon in the original 
	    spacetime~(\ref{4D_metric}) with the bifurcation
       surface replaced by a smooth surface. 

 \item[(iii)] There are the curvature singularities at 
$\tilde t=\tilde t_s(r)=-1/\tilde r$ and  $\tilde r=-1$. 
These singularities are timelike (section~\ref{sec:singularity}).
The time-dependent singularity $\tilde t_s(\tilde r)$ 
present in the $\tilde r<0 $ domain 
exists for $\tilde t>1$, whereas $\tilde t_s(\tilde r)$ 
lying in the $\tilde r>0$ region and 
the other singularity $\tilde r=-1$ 
 exist for eternity.   
\end{itemize}

These observations prompt us to imagine the positional relation
between singularities and the horizons. 
Figure~\ref{fg:fig2} describes the conformal diagram of our dynamical
black hole. 
From properties (i) and (ii), we can depict the identical 
horizon structure as in Figure~\ref{PD_static}.  
Since the $\tilde t= {\rm constant}(<\infty) $ lines are everywhere spacelike, 
each slice originates from the ``throat'' $\tilde r=0$. 
For negative values of $\tilde t$, 
$\tilde t={\rm constant}$ surfaces must intersect the singularity 
$\tilde t_s(\tilde r)$ at finite $\tilde r$ (see Figure~\ref{fig:TH}).  
Considering property (iii) that the singularity outside the horizon is only 
$\tilde t=\tilde t_s (\tilde r)<0$, the right side dotted portion of
grey line can be drawn in  Figure~\ref{fg:fig2}.   
Outside the horizon $\tilde r>0$, 
one can depict the contours of $\tilde t={\rm constant}$ and 
$\tilde r={\rm constant}$ family of surfaces, both of them to be
orthogonal~(Figure~\ref{fg:fig2}). These aspects are
all consistent with our numerical survey of geodesics.
We thus conclude that the spacetime metric~(\ref{4D_metric0})
indeed describes a black hole in the FLRW universe [aside
from the undesirable timelike naked singularity $\tilde t_s(\tilde r)$].
Although the null surface $\tilde R_-$ is a one way membrane of ``region of no entrance,''
it does not deserve to be a white hole 
horizon in a mathematical sense since the spacetime does not possess the past null infinity.

Inside the event horizon $\tilde R<\tilde R_+$,
the timelike singularities are vertically joined 
at $\tilde t=-1$. 
The past boundary $\tilde R_+$
can be matched to the black hole horizon. 
We can find as sketched in Figure~\ref{fg:fig2} that 
these patches are infinitely arrayed vertically.   
It should be emphasized, however,  that this is only a 
possible extension, since the horizon is not analytic in the present
case. One may glue the near-horizon geometry (\ref{NHmetric}) 
to the spacetime (\ref{4D_metric0}) across the horizon.

\begin{widetext}
\begin{center}
\begin{figure}[h]
\begin{tabular}{ cc }
\includegraphics[width=4.2cm]{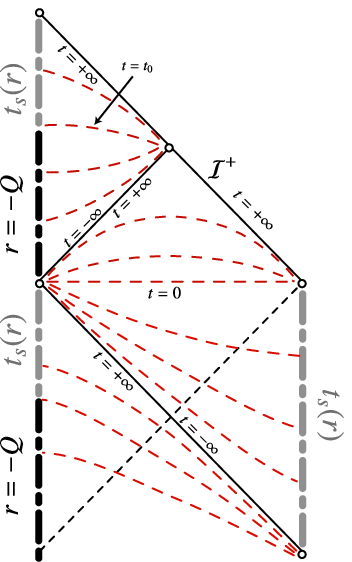}
~~~~~~~~~~~&			
\includegraphics[width=4.2cm]{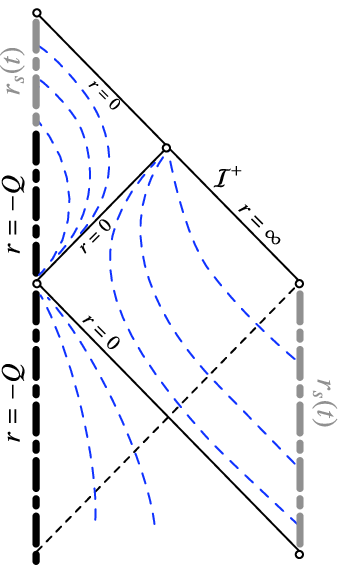} \\
(a) The contour of $\tilde t={\rm constant}$.~~~~~~~~~~~~~&
(b) The contour of $\tilde r={\rm constant}$. \\
\end{tabular}
\caption{
Conformal diagram of the spacetime~(\ref{4D_metric0}). 
We draw 
(a) the contour (spacelike) curves of $\tilde t={\rm constant}$ by red lines and  
(b) the contour (timelike) curves of $\tilde r={\rm constant}$ by blue lines. 
The singularity of $\tilde R=0$ consists of three parts: 
a black-hole singularity at $\tilde r=-1$ and 
at $\tilde r=\tilde r_s(\tilde t)=-1/\tilde t$ inside the horizon,  
and a ``big bang singularity'' 
$\tilde t=\tilde t_s(\tilde r)=-1/\tilde r$
outside the horizon. }
\label{fg:fig2} 
\end{figure}
\end{center}
\end{widetext}

Next, we wish to fill in the trapping horizons into this diagram
(we only consider the $\tau=1$ case). 
We should remind the following remarks:

\begin{itemize}
 \item[(i)] 
There are trapping horizons $\tilde t=\tilde t_{\rm TH}^{(\pm)}$ 
at which $\tilde \theta_\mp =0$. 
Outside the horizon $(\tilde r>0)$, 
the whole portion of trapping horizon $\tilde t^{(+)}_{\rm TH}$ 
is past-outer for $\tau <\tau_{\rm crit}$ hence always
spacelike, analogous to
FLRW universe filled by a stiff matter. While the   
trapping horizon $\tilde t^{(-)}_{\rm TH}$ is always timelike. 
Inside the trapping horizon ($\tilde r<0$),
$\tilde t^{(\pm)}_{\rm TH}$ coincide at $\tilde r=\tilde r_0$. 
$\tilde t^{(-)}_{\rm TH}$ and a part of $\tilde t^{(+)}_{\rm TH}$ 
near $r\simeq r_0$ are timelike. Other portion of 
$\tilde t=\tilde t_{\rm TH}^{(+)}$ changes signature near 
$\tilde r= 0$ into spacelike.

\item[(ii)] 
Trapping horizons occur where $\tilde R={\rm constant}$ surfaces
becomes null. The contour curve of circumference radius 
is spacelike for $\tilde R_2 <\tilde R<\tilde R_1$
(see Figure~\ref{fig:constantR}).  
As approaching the event horizon,  
the trapping horizons $\tilde t^{(\pm)}_{\rm TH}$ 
tend to have constant radii $\tilde R_\pm$.

\item[(iii)] 
For $\tilde r<0$, the circumference radius
$\tilde R=[(1+\tilde t\tilde r)(1+\tilde r)^3]^{1/4} $
becomes infinitely large as $\tilde t\to-\infty $
with $\tilde r$ staying constant. We can show 
following the same argument in Eq.~(\ref{distance}) that this is a 
``past timelike infinity.'' 
\footnote{
This is slightly different from 
the extremal RN geometry, for which it
takes an infinite affine time to reach
the corresponding point from inside the black hole, but
the point locates at the finite circumference radius.
The reason why the point in the present spacetime has 
an infinite circumference radius inside the horizon might be
due to the cosmic expansion.
}

\end{itemize}

\begin{figure}[t]
\includegraphics[width=4.2cm]{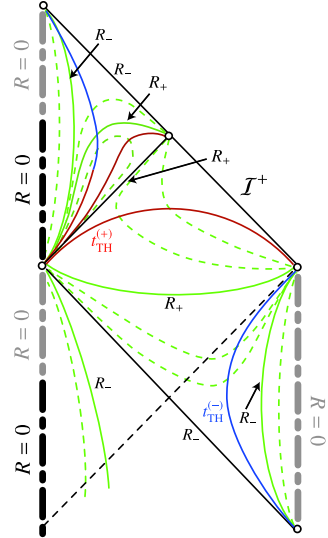} 
\caption{
A conformal diagram of the black hole in the expanding universe. 
The curves of $\tilde R={\rm constant}$ are shown by green lines. 
Null curve $\tilde R_+$  is the future event horizon,
and ${\cal I}^+$  is the future null infinity.
The trapping horizons 
$\tilde t=\tilde t^{(+)}_{\rm TH}(\tilde r)$ and 
$\tilde t=\tilde t^{(-)}_{\rm TH}(\tilde r)$
are also shown by red and blue lines.
The diagram can be extended
beyond the upper $\tilde R_-$ null curve (possibly with
the contracting patch $\tau<0$ that turns the above figure upside-down)
 in a continuous but 
a nonanalytic manner.  
} 
\label{fg:fig1} 
\end{figure}

Outside the horizon $\tilde r>0$, 
the $\tilde R={\rm constant}~(>\tilde R_+)$ surfaces
 are the same
as the FLRW cosmology: there exist a spacelike past trapping horizon
$\tilde t_{\rm TH}^{(+)}(\tilde r)$,
above which $\tilde R={\rm constant}$ surfaces are timelike and 
below which $\tilde R={\rm constant}$ surfaces are spacelike
(see Figure.~\ref{fg:fig1}). 
For $\sqrt \tau \equiv 1<\tilde R<\tilde R_+$,  $\tilde R={\rm constant}$ curves
are everywhere spacelike and lie in the future of a critical null curve
$\tilde t=\tilde t_*(\tilde r)$. 
For $\tilde R_-<\tilde R<1\equiv \sqrt \tau$,  $\tilde R={\rm constant}$ curves
cross the future trapping horizon 
$\tilde t^{(-)}_{\rm TH}(\tilde r)$ and change signature. 
For $\tilde R<\tilde R_-$, $\tilde R={\rm constant}$ curves are 
always timelike.

Inside the horizon $(\tilde r<0)$, 
$\tilde R={\rm constant}$ curves are the same as outside
for $\tilde R<\tilde R_-$ and $\sqrt \tau\equiv 1<\tilde R<\tilde R_+$. 
Whereas, $\tilde R={\rm constant}$ curves for $\tilde R>\tilde R_+$
quite differ from those in the outside. They cross the 
trapping horizons twice.

%%%%%%%%%%%%%%%%%%%%%%%%%%%%%%%%%%%%%%%%%%%%%%%%%%%%%%%%%%%
%%%%%%%%%%%%%%%%%%%%%%%%%%%%%%%%%%%%%%%%%%%%%%%%%%%%%%%%%%%
\section{Black hole thermodynamics}
%%%%%%%%%%%%%%%%%%%%%%%%%%%%%%%%%%%%%%%%%%%%%%%%%%%%%%%%%%%
%%%%%%%%%%%%%%%%%%%%%%%%%%%%%%%%%%%%%%%%%%%%%%%%%%%%%%%%%%%

Black hole thermodynamics has been established as 
rigorous mathematical laws of black 
holes with Killing horizons~\cite{BHTD1,BHTD2,Hawking1974,Wald:1975kc,Wald:1993nt,Iyer:1994ys,Iyer:1995kg,
Gao:2001ut}.  
Since the three laws of black hole thermodynamics 
interrelate the classical gravity, quantum mechanics and statistical
mechanics, they are likely to have a key r\^ole toward quantum laws of
gravity. 
Lastly, we discuss the thermodynamic properties of 
the present time-dependent black hole.

Since the present spacetime~(\ref{4D_metric0}) possesses a Killing
horizon, the thermodynamic laws continue to hold without change. 
It turned out that the Killing horizon has nonvanishing 
surface gravities~(\ref{kappa_pm}). However, their values are dependent on the
normalization of the Killing field generators of the horizon. 
Here, we intend to obtain the temperature associated to the 
time translation in the FLRW universe. 
To this end, we resort to the laws of trapping horizons. This  
is a hotly-discussed issue in recent 
years~\cite{Hayward1993,Hayward:1997jp,nozawa}.  
Laws of trapping horizons are local extensions of 
black hole thermodynamics. 

When discussing the 
dynamical aspects of black holes, 
a major obstacle for extending the black hole thermodynamics
to a non-stationary setting is how to define a
surface gravity. In the case of a spherically symmetric spacetime, 
a natural time direction is specified by the {\it Kodama
flow}~\cite{Kodama,Hayward:1997jp,nozawa}.

Write the spherically symmetric metric as, 
\begin{align}
 \D s^2 =g_{AB}(x)\D x^A \D x^B+R^2 (x)\D \Omega_2^2\,, 
\end{align}
where $g_{AB}(x)~(A, B=1,2)$ is the two-dimensional Lorentz manifold
$(M^2, g_{AB})$ perpendicular to
the metric sphere. The coordinate, $x^A$, on $M^2$ corresponds to
$t$ and $r$.   Using this coordinate patch, 
the Kodama vector is defined by~\cite{Kodama}
%---------- Kodama vector ----------%
\begin{align}
 K^A =-\epsilon ^{AB}D _B R,
\label{kodama}
\end{align}
where $\epsilon_{AB}$ and $D_A$ are the volume element and  
the covariant derivative of $g_{AB}$. We may view  $K^A$ 
as a spacetime vector by $K^\mu =K^A(\partial_A)^\mu $. 
It follows immediately from the orthogonal property
$K^\mu \nabla_\mu R=0$ that $K^\mu $ is divergence-free,
$\nabla_\mu K^\mu =R^{-2}D_A(R^{-2}K^A)=0$. 
Another key property comes from the relation
$K^\mu K_\mu =-(\nabla _\mu R)(\nabla^\mu R)$, so that $K^\mu $ 
is timelike (spacelike)
in the untrapped (trapped) region, i.e., $K^\mu $ defines a 
preferred timelike direction in the untrapped region, 
irrespective of the non-stationarity of spacetime.
Specifically, the Kodama vector becomes null 
at the trapping horizon, just as in the same way the Killing vector
becomes null at the Killing horizon.

It is enlightening here to look into the relation between $K^\mu $ 
and the Misner-Sharp energy, which is also characteristic to spherically
symmetric spacetimes~\cite{hideki}. 
Inspecting 
$\nabla_\mu K_\nu =D_A K_B(\nabla_\mu x^A)(\nabla_\nu x^B)$, 
a simple calculation shows
that ${\cal G}_{\mu \nu }\nabla^\mu K^\nu =0$ 
holds in any spherical spacetimes. 
Hence we can define a divergence-free vector field 
$\kappa ^2 J^\mu =-{\cal G^\mu }_\nu K^\nu $, representing an 
energy current due to the Einstein equations. 
Integration of $J^\mu $ over the volume $V$ with exterior boundary $S$ 
yields the Misner-Sharp energy $m=-\int_V J^\mu \D \Sigma_\mu  $.
To summarize, the Misner-Sharp energy is a charge associated with 
the locally conserved current.  

As seen above, the Kodama vector in the spherical spacetime 
plays a r\^ole similar to the Killing field in stationary spacetime.  
One can speculate that 
laws of trapping horizons are related to an observer along the Kodama
flow.

%%%%%%%%%%%%%%%%%%%%%%%%%%%%%%%%%%%%%%%%%%%%%%%%%%%%%%%%%%%
%%%%%%%%%%%%%%%%%%%%%%%%%%%%%%%%%%%%%%%%%%%%%%%%%%%%%%%%%%%
\subsection{Temperature: 0th law}
%%%%%%%%%%%%%%%%%%%%%%%%%%%%%%%%%%%%%%%%%%%%%%%%%%%%%%%%%%%
%%%%%%%%%%%%%%%%%%%%%%%%%%%%%%%%%%%%%%%%%%%%%%%%%%%%%%%%%%%

A na\"ive definition of the surface gravity for 
the trapping horizon is to replace the Killing field
by the Kodama vector in the definition of surface gravity of a Killing
horizon. This prescription does not work, since the trapping horizon is
not the null surface generated by the Kodama vector. 
A proposed definition of surface gravity for the trapping 
horizon~\cite{Hayward:1997jp,nozawa} 
is given  by the ``equilibrium part''
\begin{align}
K^\nu \nabla_{[\nu }K_{\mu ]}=\kappa _{\rm TH} K_\mu
\,,
\end{align}
where the equality is evaluated at the trapping horizon. 
After some amount of algebra, one finds that the 
surface gravities of trapping horizons $\tilde t=\tilde t^{(\pm )}_{\rm TH}$ 
are given by 
\begin{align}
\tilde \kappa^{(+)}_{\rm TH} = 
\frac{(1+\tilde r)^6(1+4\tilde r)^3}{8\tilde
 R_1^9}\left[\frac{4\tilde r^2+8 \tilde r+1}{(1+\tilde r)^8}
\tilde R_1^8-1\right]
\,, \\
\tilde \kappa^{(-)}_{\rm TH} = 
\frac{(1+\tilde r)^6(1+4\tilde r)^3}{8\tilde
 R_2^9}\left[\frac{4\tilde r^2+8\tilde r+1}{(1+\tilde r)^8}\tilde R_2^8-1\right]
\,.
\end{align}
Taking the event horizon limit $\tilde r \to 0$ 
[see Eqs.~(\ref{r0limit1}) and (\ref{r0limit2})] in
the above equations,  
we obtain the black hole temperature 
\begin{align}
T_{\rm BH}^{(\pm )}:=&\left.{\tilde \kappa
 ^{(\pm)}_{\rm TH} \over 2\pi Q}\right|_{\tilde r\to 0}
={\sqrt{1+4\tau^2}\over 16\pi\tau^2 \tilde R_\pm^5 Q}\,.
\label{temperature}
\end{align}
Comparing this with Eq.~(\ref{kappa_pm}), 
these are equivalent to surface gravities associated with 
renormalized generator of the horizon,  
\begin{align}
 \xi ^\mu =\left(\frac{\partial }{\partial \tilde T_\pm}\right)^\mu \
 \to \ 
\frac{1}{4\tau \tilde R_\pm^3} 
\left(\frac{\partial }{\partial \tilde T_\pm}\right)^\mu
 \,,
\end{align} 
which coincides with the Kodama vector  evaluated on the horizon
 for the near-horizon metric~(\ref{NHmetric2}).

For the future horizon $\tilde R_+$, the temperature 
$T_{\rm BH}^{~(+)}$ takes the maximum value 
\begin{eqnarray}
T_{\rm BH (max)}^{(+)}={3^{-{9\over 4}}\over  2\pi Q}\approx 0.00213856 
\, Q^{-1}
\end{eqnarray}
 at $\tau=\sqrt{3}/2$ for fixed charge.
The temperature vanishes in both limits of
$\tau\rightarrow \infty$ (degenerate horizon) 
and $\tau\rightarrow 0$ (no horizon).
The former recovers the result for 
the extremal RN black hole. 
The temperature at the past horizon $\tilde R_-$, on the other hand, 
has no maximum value.
It monotonically increases to infinity as
$\tau\rightarrow 0$.
$T_{\rm BH}^{(-)}$ is always higher than $T_{\rm BH}^{(+)}$.

\subsection{Energy balance: 1st law}

It is a widely accepted criterion that 
a well-defined energy should satisfy an energy balance law.
The Misner-Sharp energy indeed fulfills this, 
as in Eq.~(\ref{variationFormula}). 
We can rewrite each term in this equation into more
recognizable form. Defining 2D quantities,
\begin{align}
P_{\rm eff}=-\frac 12 {T^A}_A, \qquad
\psi_A = T_{AB}D^B R+PD_A R\,,
\end{align}
and making use of the Einstein equations, 
one arrives at the {\it unified first law}~\cite{Hayward:1997jp},
%------------- unified 1st law  ---------------%
\begin{align}
D_A m =A\psi_A+P_{\rm eff}D_A V\,,
\label{unfied1stlaw}
\end{align}
where $A=4\pi R^2$ and $V=\frac{4\pi}3R^3$ denote the 
area and  volume of the metric sphere. 
This equation illustrates that the mass variation is supplied by 
an injection of energy current and the external work term. 
The expression of $\psi_A$ is comparatively messy, but it 
is straightforward to obtain.

Projecting Eq.~(\ref{unfied1stlaw})
along the generator , $\zeta ^\mu =\zeta^A (\partial_A)^\mu $, 
of the trapping horizon and noticing the fact that  
$\zeta^\mu \nabla_\mu (m/R)=0$, one obtains 
%------------- 1st law of TH  -------------%
\begin{align}
A \zeta^A \psi_A =\frac{\kappa _{\rm TH}}{8\pi} \zeta^A D_A A\,. 
\end{align} 
This is an energy balance law of a trapping horizon (see
\cite{Hayward:1997jp,hideki} for detailed derivation). 
Here, along the trapping horizon with $\theta_+=0$, 
$\zeta^A$ is obtained as 
$\zeta^A=\pm \epsilon ^{AB}D_B\theta _+$ (where the sign should
be appropriately chosen in such a way that it is outgoing in the spacelike
case or future-directed in the timelike case).

Using Einstein's equations, we find that the surface gravity is 
expressed in terms of the Misner-Sharp energy and the pressure as
\begin{align}
\kappa_{\rm TH} =\frac{m}{R^2} -4\pi R P_{\rm eff}\,,
\end{align}
where equality is understood at the trapping horizon.
This is the Newtonian analogue definition of acceleration.

%%%%%%%%%%%%%%%%%%%%%%%%%%%%%%%%%%%%%%%%%%%%%%%%%%%%%%%%%%%
%%%%%%%%%%%%%%%%%%%%%%%%%%%%%%%%%%%%%%%%%%%%%%%%%%%%%%%%%%%
\subsection{Entropy: 2nd law}
%%%%%%%%%%%%%%%%%%%%%%%%%%%%%%%%%%%%%%%%%%%%%%%%%%%%%%%%%%%
%%%%%%%%%%%%%%%%%%%%%%%%%%%%%%%%%%%%%%%%%%%%%%%%%%%%%%%%%%%

It follows from the first law of a trapping horizon that 
we can identify the 
 entropy  by one quarter of the area of the trapping horizon,
i.e., 
it accords with the ``Bekenstein-Hawking
formula,''
\begin{align}
 S_{\rm TH}^{(+)}=\frac{A_1}{4G}=\frac{\pi R_1^2}{G}\,, \qquad
 S_{\rm TH}^{(-)}=\frac{A_2}{4G}=\frac{\pi R_2^2}{G}.
\end{align}
Taking the event horizon limit~(\ref{r0limit1}) or (\ref{r0limit2}), 
we recover the well known result~\cite{Hawking1974,Wald:1993nt} 
\begin{align}
S_{\rm BH}^{(\pm)}={{A}_{\rm BH}^{(\pm )}\over 4G}=
{\pi Q^2 \over 2G\tau}\left(\pm 1+\sqrt{1+4\tau^2}\right)
\,.
\end{align}
In the limit $\tau\to \infty $ with fixed charge,  
the above entropy reduces to that of the extremal RN
black hole, $S_{\rm BH}^{(\pm)}={\pi Q^2 / G}$.

%%%%%%%%%%%%%%%%%%%%%%%%%%%%%%%%%%%%%%%%%%%%%%%%%%%%%%%%%%%
\section{Concluding Remarks}
\label{conclusion}
%%%%%%%%%%%%%%%%%%%%%%%%%%%%%%%%%%%%%%%%%%%%%%%%%%%%%%%%%%%

In this paper, we have made a thorough discussion about the 
causal structure and physical properties of the spacetime
derived from intersecting M-branes. 
We have found that the solution indeed describes a black hole
embedded in the FLRW cosmology filled with fluid obeying the stiff
equation of state. The global causal structure is displayed in 
Figures~\ref{fg:fig2} and \ref{fg:fig1}. 
Since the solution is approximated by the extreme RN solution near the ``throat'' 
and the flat FLRW universe with $P=\rho $ at infinity, one might first
envisage that the causal structure is obtainable by patching these two
limiting spacetimes. That is to say, according to our first intuition, 
one might have expected that 
spacetime should possess a spacelike big-bang singularity at $t=0$, 
there should exist a degenerate event horizon, and the timelike
singularity should appear only inside the hole. 
However, our careful analysis revealed that the global causal structure 
is completely different from the above rough estimate.

Our solution satisfies the dominant energy condition, so that 
the energy densities are always positive, absolute value of 
principal pressures do not exceed the energy density  for respective
fluids and  the energy flux current is always causal. 
This desirable property is not seen in the solutions found in the
literature. Hence the results presented in this paper open up new
avenues for further research on black holes surrounded by  
usual matters in the expanding universe from higher-dimensional 
point of view. 
Our solution, however, may not have a direct astrophysical relevance 
because of nonzero electromagnetic charge.  
The charge is probably also responsible for the timelike singularity
$t=t_s(r)$ outside the horizon.  The timelike singularity does not
develop as the big-bang singularity in the usual 
FLRW cosmology with fluid $P=w \rho ~(-1\le w\le 1)$.  
Unfortunately, the construction of black hole solution without charge 
may be beyond the intersecting brane picture.

In the process for obtaining the global structure, we gave a coherent
description concerning the trapping horizons. The main idea on which
our discussion is based is that the trapping horizons reflect
the physical situations of marginal surfaces on which the either of 
expansions of light ray vanishes. 
This local character enables us to relate it to the curvature singularity and 
the Misner-Sharp energy. A more important belief to which we resort is
that the trapping horizon with negative outgoing expansion does not
occur outside the horizon. The present spacetime indeed has this
property (except in the neighborhood of singularity). 
We confirm the infinite redshift (blueshift) surface as a black hole
horizon (white hole horizon in a quoted sense) 
combining the analysis of near-horizon geometry and the 
behaviors of null geodesics. 

It was somewhat surprising that the solution admits a nondegenerate Killing
horizon. The Killing horizon is usually associated with
symmetry of time-translation and angular-rotation. 
The black hole remains the same size and fails to grow, although the 
the black hole is surrounded with fluid. This
characteristic property may be ascribed to the fact that the 11D
solution was supersymmetric in the static limit. Although the
dynamically brane intersecting solution breaks supersymmetry, 
it still maintains a part of the BPS characters.   
The same takes place in the Kastor-Traschen black hole.

In this paper, we have taken a particular notice on 
the solution, whose 11D ``oxidized'' solution has four kinds of 
harmonics of spherical symmetry. A more general non-spherical 
spacetime is of course more complicated. 
Still, the  profound understanding of
spherically symmetric case will of substantial aid in 
exposing  more complex structures of dynamical black holes.

%%%%%%%%%%%%%%%%%%%%%%%%%%%%%%%%%%%%%%%%%%%%%%%%%%%%%%%%%%%
%%%%%%%%%%%%%%%%%%%%%%%%%%%%%%%%%%%%%%%%%%%%%%%%%%%%%%%%%%%
%%%%%%%%%%%%%%%%%%%%%%%%%%%%%%%%%%%%%%%%%%%%%%%%%%%%%%%%%%%
\acknowledgments
%%%%%%%%%%%%%%%%%%%%%%%%%%%%%%%%%%%%%%
%%%%%%%%%%%%%%%%%%%%%%%%%%%%%%%%%%%%%%

We would like to thank Gary W. Gibbons, Hideki Maeda, Nobuyoshi 
Ohta and Harvey S. Reall 
for valuable  comments and discussions.
KM would acknowledge hospitality of 
DAMTP and the Centre for Theoretical Cosmology,
Cambridge University
during his stay in September, 2009.
This work was partially supported 
by the Grant-in-Aid for Scientific Research
Fund of the JSPS (No.19540308) and for the
Japan-U.K. Joint Research Project,
and by the Waseda University Grants for Special Research Projects.

%%%%%%%%%%%%%%%%%%%%%%%%%%%%%%%%%%%%%%%%%%%%%%%%%%%%%%%%%%%
%%%%%%%%%%%%%%%%%%%%%%%%%%%%%%%%%%%%%%%%%%%%%%%%%%%%%%%%%%%
\appendix
%%%%%%%%%%%%%%%%%%%%%%%%%%%%%%%%%%%%%%%%%%%%%%%%%%%%%%%%%%%
\section{Intersecting brane and black holes}
\label{Intersecting_brane}
%%%%%%%%%%%%%%%%%%%%%%%%%%%%%%%%%%%%%%%%%%%%%%%%%%%%%%%%%%%

In this Appendix, we consider an intersecting brane system
in the 11D supergravity theory, which is expected to be an effective
field theory of M-theory. We discuss how to obtain the 4D effective action 
and produce solutions in the 4D spacetime. We intend to consider
M-branes, for which the Chern-Simons term  
`` $F \wedge F \wedge A $'' has no contribution. Hence, it suffices 
in our setup to concentrate on the following 11D effective action, 
\begin{widetext}
\begin{eqnarray}
{\cal S}={1\over 2\kappa_{11}^2}\int \D ^{11}X
\sqrt{-g_{11}}\left[
{\cal R}_{11}-\sum_A{1\over 2(p_A+2)!}({\cal F}_{p_A+2})^2\right]
\label{11D_action}
\,.
\end{eqnarray}
Here, $A$ denotes the type of branes with which the Abelian ($p_A+2$)-form field
 $F_{p_A+2}$ is coupled, and $p_A $(=$2$ or $5$) is the dimensions of branes. 

Once the 11D brane configuration is given, the 4D solution is derivable 
via the standard toroidal compactification. 
We shall analyze intersecting brane systems involving four-charges, which is needed 
to find a 4D maximally charged (supersymmetric or nonsupersymmetric) 
black hole with regular event horizon. 
We can construct two kinds of such a configuration: M2-M2-M5-M5 and
M2-M5-W-KK system. As a concrete example, we focus our attention to the 
 M2-M2-M5-M5 intersecting brane system. Compatibility with  11D
 supergravity equations of motion determines 
respective brane codimensions, which is given by Table~\ref{table1}.

\begin{table}[h]
{\normalsize
\begin{center}
\begin{tabular}{|c|c|c|c|c|c|c|c|c|c|c|c|}
\hline
& $0$ & $1$ & $2$ & $3$ &$4$
&$5$& $6$ & $7$ &$8$ &$9$ &$10$
\\
\hline
M5 & $\circ$ & $\circ$ & $\circ$ & $\circ$ & $\circ$ & $\circ$ &&&&&
\\
\cline{2-12}
M5 & $\circ$ & $\circ$ & $\circ$ & $\circ$ &&& $\circ$ & $\circ$ & & &
\\
\cline{2-12}
M2 & $\circ$ &  &  &   &$\circ$&& $\circ$ & &  & &
\\
\cline{2-12}
M2 & $\circ$ &  &   &  &&$\circ$ & & $\circ$ & & &
\\
\hline
\end{tabular}
\caption{M2-M2-M5-M5 brane system. The circles describe 
which dimensions are filled by the corresponding branes.}
\label{table1}
\end{center}
}
\end{table}

There appear four charges ($Q_2, Q_{2'}, Q_5, Q_{5'}$) associated 
with the corresponding 
four branes. In the static spacetime, we have the following 
intersecting  brane solution: 
\begin{eqnarray}
\D s^2&=&H_2^{1/3}H_{2'}^{1/3}H_5^{2/3}H_{5'}^{2/3}
\left[
-H_2^{-1}H_{2'}^{-1}H_5^{-1}H_{5'}^{-1}\D t^2+
H_5^{-1}H_{5'}^{-1}
\left(
\D y_1^2+\D y_2^2+\D y_3^2
\right)
\right.
\nonumber \\
&&
\left.
+H_5^{-1}H_{2}^{-1}\D y_4^2+H_5^{-1}H_{2'}^{-1}\D
y_5^2+H_2^{-1}H_{5'}^{-1}\D y_6^2
+H_{2'}^{-1}H_{5'}^{-1}\D y_7^2
+(\D r^2+r^2\D \Omega^2_2)
\right]
\label{11D_metric0}
\,,
\end{eqnarray}
\end{widetext}
where $H_A$ are harmonics on the three Euclidean space 
($\D s_3^2=\D r^2+r^2 \D \Omega^2_2$).  One can see immediately that 
the directions involving inverse $H_A$ for the metric in the square bracket  
correspond to the dimensions to which the $A$-brane belong.

If one toroidally compactifies common 7D world-volume of branes, 
a 4D solution is obtained. 
Rewriting the 11D metric as
\begin{eqnarray}
\D s^2=\prod_{i=1}^7 b_i^{-1} \times \D s_4^2+\sum_{i=1}^7 b_i^2 \D y_i^2
\label{11D_metric}
\,,
\end{eqnarray}
where
\begin{eqnarray}
&&
b_1^2=b_2^2=b_3^2=\left({H_2H_{2'}\over H_5H_{5'}}\right)^{1/3}
\,,
\end{eqnarray}
\begin{eqnarray}
&&
b_4^2=\left({H_{2'}H_{5'}^2\over H_2^2H_5}\right)^{1/3}
\,,~
b_5^2=\left({H_2H_{5'}^2\over H_{2'}^2H_5}\right)^{1/3}
\,,~
\nonumber
\\
&&
b_6^2=\left({H_{2'}H_5^2\over H_2^2H_{5'}}\right)^{1/3}
\,,~
b_7^2=\left({H_2H_5^2\over H_{2'}^2H_{5'}}\right)^{1/3}
\,,~~~~
\end{eqnarray}
and compactifying $y_i$-coordinates ($i=1, \cdots, 7$),
the 4D solution in the Einstein-frame is given by 
\begin{eqnarray}
\D s_4^2=-\Xi \D t^2+{1\over \Xi}\left(\D r^2+r^2\D \Omega^2_2\right)\,,
\label{4D_metric20}
\end{eqnarray}
where 
\begin{eqnarray}
\Xi=\left(H_2H_{2'}H_5H_{5'}\right)^{-1/2}
\,.
\end{eqnarray}

If we assume that 
 harmonics $H_A$'s are spherically symmetric,
i.e., 
\begin{align}
&
H_2=1+{Q_2\over r}\,,\qquad 
H_5=1+{Q_5\over r}\,,
\nonumber \\
&
H_{2'}=1+{Q_{2'}\over r}\,,\qquad 
H_{5'}=1+{Q_{5'}\over r}\,.
\label{static}
\,
\end{align}
we find a static extreme black hole solution in four dimensions.
Here, $Q_A$'s represent the brane charges. 
If all charges vanish, both of the 
11D and 4D solutions are trivial.
The extension of the harmonic functions $H_A$ as 
discussed in Appendix \ref{multi_BH} gives a multi-black hole
system.

Generalizing this solution to the time-dependent one,
we find that only one brane among four can be time-dependent 
under the metric ansatz assumed in~\cite{MOU}. 
The intersecting brane metric is still given by Eq.~(\ref{11D_metric0}).  
The field equations require that the time-dependence is linear, i.e., 
the metric functions in the spherically symmetric case are 
given by
\begin{align}
&
H_T={t\over t_0}+{Q_T\over r}\,,\qquad
H_S=1+{Q_S\over r}\,,
\nonumber \\
&
H_{S'}=1+{Q_{S'}\over r}\,,\qquad 
H_{S''}=1+{Q_{S''}\over r}\,,
~~~~
\label{time_dependent}
\end{align}
where $t_0$ is a constant with dimension of time.
$T$ and $S, S', S''$ denote one time-dependent brane and
three static branes, respectively. Any one of
M2, M2$'$, M5, and M5$'$ branes can have time-dependence.
This gives a black hole in the expanding universe in four dimensions, 
which we discuss in this paper.
It is also extended to a multi-black hole system (see Appendix
\ref{multi_BH}).
If all brane charges are set to zero,
11D solution is the Kasner solution describing a homogeneous but anisotropic
vacuum universe, 
whereas the 4D solution reduces to the flat FLRW cosmology.

Since we know the 11D action (\ref{11D_action}), 
assuming the brane configuration shown in Table~\ref{table1} 
and compactifying the spatial directions 
as (\ref{11D_metric}), we can derive
the effective 4D action, which gives 
the present time-dependent solutions, as follows.

The scales of extra dimensions $b_i$ ($i=1, \cdots, 7$) behave as
scalar fields in 4D spacetime, i.e.,
the effective action of gravity sector is
\begin{widetext}
\begin{eqnarray}
{\cal S}_4(g,b_i)=\int \D ^{4}x
\sqrt{-g}\left\{
{1\over 2\kappa^2}R-{1\over 4}\left[
 \left(\nabla\sum_{i=1}^7
\ln b_i\right)^2+2\sum_{i=1}^7\left(\nabla\ln b_i\right)^2
\right]
\right\}
\,.
\end{eqnarray}
 
Although we compactify seven dimensions, we have only four branes.
Hence the degrees of freedom are maximally four. 
This can be affirmed by 
writing down the kinetic term of the scalar fields
in terms of harmonic functions $H_A$ as
\begin{eqnarray}
&&{1\over 4}\left[
 \left(\nabla\sum_{i=1}^7
\ln b_i\right)^2+2\sum_{i=1}^7\left(\nabla\ln b_i\right)^2
\right]
\nonumber \\
&&
={1\over 16 }
\left[3\left\{(\nabla \ln H_2)^2+(\nabla \ln H_{2'})^2+(\nabla \ln H_5)^2
+(\nabla \ln H_{5'})^2\right\}
-2\left(\nabla \ln H_2\cdot\nabla \ln H_{2'}
+\nabla \ln H_5\cdot\nabla \ln H_{5'}\right)
\right.
\nonumber \\
&&
\left.
~~~-2\left(\nabla \ln H_2\cdot\nabla \ln H_{5}
+\nabla \ln H_2\cdot\nabla \ln H_{5'}+\nabla \ln H_{2'}\cdot\nabla \ln H_{5}
+\nabla \ln H_{2'}\cdot\nabla \ln H_{5'}\right)
\right]
\nonumber \\
&&
={1\over 16}
\left[
(\nabla \ln (H_2/H_{2'}))^2+(\nabla \ln (H_{5}/H_{5'}))^2
+(\nabla \ln (H_2/H_5))^2
\right.
\nonumber\\
&&
\left.
~~~+(\nabla \ln (H_2/H_{5'}))^2
+(\nabla \ln (H_{2'}/H_{5}))^2+(\nabla \ln (H_{2'}/H_{5'}))^2
\right]
\nonumber \\
&&={\kappa^2\over 2}\sum_{A<B}(\nabla \phi_{AB})^2
\,.
\label{kinetic_phiAB}
\end{eqnarray}
\end{widetext}
where 
\begin{eqnarray}
\kappa \phi_{AB}={1\over 2\sqrt{2}}\ln \left({H_A\over H_{B}}\right)
\,,
\label{phiAB}
\end{eqnarray}
denotes the ``scalar field mixing'' term.

Supposed that 
all charges are equal ($Q_2=Q_{2'}=Q_5=Q_{5'} \equiv Q$) as in the main text,
it follows that  
$\phi_{SS'}, \phi_{S'S'\hspace{-.1em}'}, \phi_{SS'\hspace{-.1em}'}$ are 
trivial, and $\phi_{TS}, \phi_{TS'}, \phi_{TS'\hspace{-.1em}'}$ are the same.
As a result only a single scalar field $\Phi$ survives, 
which is normalized from (\ref{kinetic_phiAB}) as

\vskip -1em
\begin{eqnarray}
\kappa \Phi= \sqrt{3}\kappa\phi_{TS}
={\sqrt{3}\over 2\sqrt{2}}\ln \left({H_T\over H_S}\right)
\,.
\nonumber
\end{eqnarray}
This is identical to Eq.~(\ref{Phi}). 

Next we reduce the 11D form-field sector as follows:
the M2 and M5 branes couple to four-form and its dual
seven-form field, respectively.
\begin{widetext}
Hence the effective action of the form fields
is reduced to four dimensions as
\begin{eqnarray}
&&{1\over 2\kappa^2}\times 
\sqrt{-g_{11}}\sum_A{1\over 2(p_A+2)!}({\cal F}_{p_A+2})^2
=\left(\prod_i b_i^{-1}\right)^2\prod_i b_i\sqrt{-g}\times
\left(\prod_i b_i^{-1}\right)^{-2}{g^{\mu\rho}g^{\nu\sigma}\over 8\pi}
\nonumber \\
&&
\times
\left[
F_{\mu\nu}^{(2)}
F_{\rho\sigma}^{(2)}b_4^{-2}b_6^{-2}
+F_{\mu\nu}^{(2')}
F_{\rho\sigma}^{(2')}b_5^{-2}b_7^{-2}
+F_{\mu\nu}^{(5)}
F_{\rho\sigma}^{(5)}b_1^{-2}b_2^{-2}b_3^{-2}b_4^{-2}b_5^{-2}
+F_{\mu\nu}^{(5')}
F_{\rho\sigma}^{(5')}b_1^{-2}b_2^{-2}b_3^{-2}b_6^{-2}b_7^{-2}
\right]~~~~~~
\nonumber \\
&&
=\sqrt{-g}\times {1\over 4}\left[
\left({H_T\over H_S}\right)^{3/2}(F_{\mu\nu}^{(T)})^2
+3\left({H_T\over H_S}\right)^{-1/2}(F_{\mu\nu}^{(S)})^2\right]
\,,
\end{eqnarray}
\end{widetext}
where we set 
\begin{eqnarray}
(F_{\mu\nu}^{(A)})^2={2\pi \over \kappa^2 (p_A+2)!}({\cal F}_{p_A+2})^2\,.
\end{eqnarray}
This ansatz is consistent with our result (\ref{U(1)_potential}) 
for the Maxwell fields, because 
the electric potential of four-form field in 11D
is given by 
${\cal A}_0^{(A)}=1/H_A+a^{(A)}(t)$,
where $a^{(A)}(t)$ is an arbitrary function of $t$, which 
comes from a
gauge freedom.
As a result, we  obtain
the effective 4D action for the form-field sector as
\begin{align}
{\cal S}_4(F)=&{1\over 16\pi}\int \D ^{4}x
\sqrt{-g}\left[
e^{\mp\sqrt{6}\kappa\Phi}(F_{\mu\nu}^{(T)})^2\right. \nonumber \\&
\left.+3e^{\pm{\sqrt{6}}\kappa\Phi/ 3}(F_{\mu\nu}^{(S)})^2\right]
\,,
\end{align}
which is the same as Eq.~(\ref{4D_action}).

In the static case with equal charges, 
the 4D solution~(\ref{4D_metric20}) with (\ref{static})
corresponds to an extreme RN black hole (which is indeed a
solution in the Einstein-Maxwell system).  
While for the time-dependent case with equal charges (\ref{time_dependent}),
it describes a black hole in the FLRW universe 
which we have established in the body of the present paper.

%%%%%%%%%%%%%%%%%%%%%%%%%%%%%%%%%%%%%%%%%%%%%%%%%%%%%%%%%%%
\section{Multi black holes in the time-dependent universe}
\label{multi_BH}
%%%%%%%%%%%%%%%%%%%%%%%%%%%%%%%%%%%%%%%%%%%%%%%%%%%%%%%%%%%

Writing $H_T =t/t_0+\bar H_T$, 
$H_S=1+\bar H_S$ and so on, 
the 11D supergravity equations of motion 
require the functions $\bar H_A$  $(A=T, S, S',S'')$
to be arbitrary harmonics on flat three space. Hence, 
just by replacing the monopole term $Q_A/r$ 
by multicenter harmonics, we obtain a 
collection of black holes in a dynamical background. 
To be specific, we have
\begin{align}
\bar H_A=  \sum_i ^N\frac{Q^{(A)}_i}{|\,\vect{r}-\,\vect{r}_i^{(A)}|}\,,
\label{multi_HA}
\end{align}
where the constants $\,\vect{r}_i^{(A)}$ and
$Q^{(A)}_i ~(>0)$ correspond to the loci and the charges of $i$-th black hole 
associated with $A$-branes, respectively. 
The linear term $(\,\vect{a}_i\cdot \,\vect{r}_i)$ has been dropped
by the asymptotic boundary conditions at infinity. 
In the case of three equal harmonics 
$\bar H_S=\bar H_{S'}=\bar H_{S''}$, 
this metric solves the 
field equations of Einstein-Maxwell-dilaton system (\ref{4D_action})
if the dilaton and U(1)-gauge potentials are given by
 (\ref{Phi}) and  (\ref{U(1)_potential}).

Near each mass point (with $t$ being finite) $\,\vect{r}^{(A)}_i$,
there exists an infinite throat as in the single mass case 
discussed in the body of text. Far from the throat, on the other hand, 
the solution tends to an FLRW universe filled by a stiff matter. 

As in the case of the Kastor-Traschen solution~\cite{KT,BHKT}, 
this spacetime is expected to describe a collision of black holes
provided the background universe is contracting ($t_0<0$). 
Since each black hole is ignorant of others, i.e., 
the gravitational and electromagnetic forces between black holes are balanced, 
the collision occurs by a brute-force method responsible for 
the background contracting universe. The difference from the
Kastor-Traschen case 
lies in that the background universe obeys the power-law contraction
$a \propto \bar t^{1/3}$, so our discussion
parallels~\cite{Gibbons:2005rt}
in which colliding D3 branes were discussed in detail.

Let us start with the negative time $t<0$ and run time forwards.  
Since the $t=0$ surface is again nonsingular,  the universe continues to
shrink for positive values of $t$ until the singularity $H_T=0$ is reached. 
Specifically, the metric continues to exist inside the domain, $D_t$,
bounded by the level set $\bar H_T=t/(-t_0)$. 
It then follows that at small positive $t$, the domain $D_t$ 
is a large connected volume containing all black holes. 
As time passes, the domain $D_t$ continues to contract and 
tends to spilt into disconnected pieces containing each
mass point $\,\vect{r}_i^{(A)}$. This means that black holes
 scatter off rather than coalesce, and the universe is bounded by 
curvature singularity at $H_T=0$.

The multiple black-hole solution~(\ref{multi_HA}) is to be compared with the Kastor-Traschen solution, 
\begin{align}
\D s^2=-H^{-2}\D t^2+H^2 \D \,\vect{r}^2\,,
\end{align}
with $ \kappa F=\D (H^{-1})\wedge \D t$ and 
\begin{align}
 H=\frac{t}{t_0}+\bar H\,, \qquad \bar H:=\sum_i^N \frac{Q_i}{
|\,\vect{r}-\,\vect{r}_i|}\,.
\end{align}
This is an exact solution of Einstein-Maxwell-$\Lambda (\equiv 3/t_0^2)$ system.  
The distinction between  our spacetime and 
the Kastor-Traschen solution  is essentially only the power of the lapse
function, where divergence of lapse corresponds to the curvature singularity for
each solution.  
The exponent is closely associated with the number of branes and more 
general class of solutions is available. 
Further detailed analysis will be reported elsewhere~\cite{GMII}.

%%%%%%%%%%%%%%%%%%%%%%%%%%%%%%%%%%%%%%%%%%%%%%%%%%%%%%%%%%%
\section{5D time-dependent black holes}
\label{5DBH}
%%%%%%%%%%%%%%%%%%%%%%%%%%%%%%%%%%%%%%%%%%%%%%%%%%%%%%%%%%%

As proved in~\cite{MOU},  
a 5D time-dependent ``black hole solution'' is also obtained 
from the M2-M2-M2 and M2-M5-W intersecting brane systems. 
Let us discuss the former case. 
As in the
static counterparts, 
we need only three nontrivial charges to obtain a black hole solution. 
The 5D metric in the Einstein frame reads 
\begin{eqnarray}
\D s_5^2=-\Xi^2 \D t^2+\Xi^{-1}\left(\D r^2+r^2 \D \Omega_3^2\right)\,,
\label{5D_metric}
\end{eqnarray}
with 
\begin{eqnarray}
\Xi = 
\left(H_T
H_S
H_{S'}
\right)^{-1/3}
\,.
\end{eqnarray}
Here we have introduced 
\begin{eqnarray}
H_T&=&{t\over t_0}+{Q_T\over r^2}
\,,\nonumber
\\
H_S&=&1+{Q_{S}\over r^2}
\,,\nonumber
\\
H_{S'}&=&1+{Q_{S'}\over r^2}
\,.
\end{eqnarray}
to denote the harmonics in the flat 4D space. 
$Q_T$ and $ Q_{S}, Q_{S'}$ are 
charges of one time-dependent and two remaining static M2 branes,
respectively. The lapse function $\Xi $ takes a relatively
simple form compared to the 4D metric~(\ref{Xi}).

Assuming $t/t_0>0$ and transforming to the 
new time coordinate $\bar t$ given by 
\begin{eqnarray}
{\bar t \over \bar t_0}=\left({t \over t_0}\right)^{2\over 3} 
~~{\rm with}~~~
\bar t_0={3t_0\over 2}
\,,
\end{eqnarray}
the solution~(\ref{5D_metric}) is cast into the form,
\begin{eqnarray}
\D s_5^2&=&-\bar \Xi^2 \D \bar t^2+{a^2\over \bar \Xi}
\left(\D r^2+r^2 \D \Omega_2^3\right)\,,
\label{5D_metric1}
\end{eqnarray}
where 
\begin{eqnarray}
\bar \Xi&=&\left(\bar H_TH_SH_{S'}\right)^{-1/3}\,,
\\
a&=&\left({\bar t/ \bar t_0}\right)^{1/4}
\label{scale_factor_5}
\,,
\end{eqnarray}
with
\begin{eqnarray}
\bar H_T&=&1+{Q_T\over a^6 r^2}
\,.
\end{eqnarray}

The expansion law with (\ref{scale_factor_5})
is again identical to that of the 5D universe with a stiff matter.
The limit of $r\rightarrow 0$ with keeping $t$ finite
gives the same ``throat'' geometry of the 5D extreme RN black hole. 
According to the detailed argument laid out in the main text, 
we may regard this solution as 
a black hole in the expanding universe.

For the case in which $Q_{T}=Q_S=Q_{S'}=\equiv Q$, 
the 5D metric is an exact solution of 
Einstein-Maxwell-dilaton system 
whose action is given by
\begin{eqnarray}
{\cal S}&=&\int \D ^4x\sqrt{-g}\left[
{1\over 2\kappa^2}{\cal R}-{1\over 2}(\nabla\Phi)^2
\right.
\nonumber \\
&&~~~~~~~~~
\left.
-{1\over 16\pi}\sum_{A} e^{\lambda_A\kappa\Phi}
(F_{\mu\nu}^{(A)})^2\right]
\,,
\label{5D_action}
\end{eqnarray}
if 
the dilaton is given by
\begin{align}
\Phi = \frac{1}{\sqrt 3} \ln \left(
\frac{H_T}{H_S}\right), 
\end{align}
and the electromagnetic fields take the form, 
\begin{align}
\kappa A_{0}^{(T)}&=\sqrt{2\pi}\frac{1}{H_T}\,,
\nonumber \\
\kappa A_{0}^{(S)}&=\kappa A_{0}^{(S')}\,=\,
\sqrt{2\pi}\left(\frac{1}{H_S}-1\right)\,,
\end{align}
with coupling constants
\begin{align}
\lambda _T=\frac{4}{\sqrt 3}, \qquad 
\lambda _S=\lambda _{S'}=-\frac{2}{\sqrt 3}.
\end{align}

After short calculations, we find the following results.
The horizon radii are given by 
\begin{eqnarray}
R_\pm^3={ Q^{3/2} \over \tau}\left(\sqrt{1+16\tau^2}\pm 1\right)
\,,
\end{eqnarray}
where
\begin{eqnarray}
\tau^2={t_0^2\over Q}
\,.
\end{eqnarray}
$R_+$ and $R_-$ correspond to the future event horizon and
the past event horizon, respectively.
The Carter-Penrose diagram is quite similar to that in 4D,
although there exist minor differences.

The surface gravities are found to be
\begin{eqnarray}
\kappa_{\rm BH}^{(\pm)}={\sqrt{1+16\tau^2}\over 12\tau^2 Q^{1/2}\tilde R_\pm^7}
\,.
\end{eqnarray}
The temperature of the future event horizon 
[$T_{\rm TH}^{(\pm)}=\kappa_{\rm BH}^{(\pm)}/(2\pi)$]
vanishes in the both 
limits of $\tau\rightarrow 0$ and of $\tau\rightarrow \infty$,
just as the 4D black hole.
The maximum temperature is given by\\
 
\begin{align}
T_{\rm BH (max)}^{(+)}&={4\sqrt[12]{13}\sqrt[3]{1+\sqrt{13}}\sqrt{62+14\sqrt{13}}\over 
\pi\left(6+\sqrt{62+14\sqrt{13}}\right)^{7/3}{Q}^{1/2}}\,
\nonumber
\\
&
\approx 0.0395465\,Q^{-1/2}\,,
\end{align}
at \begin{eqnarray}
\tau={\sqrt[4]{13}\left(\sqrt{13}+1\right)\over 24}\approx 0.364381
\,.
\end{eqnarray}

%--------------------------------------------%
%                                            %
%                  References                %
%                                            %
%--------------------------------------------%

\end{document}